\documentclass{aa}

\usepackage{physics}
\usepackage{graphicx}
\usepackage{placeins}
\usepackage{verbatim}
\usepackage{ulem}
\usepackage[colorlinks=true, linkcolor=blue, citecolor=blue, filecolor=blue, urlcolor=blue]{hyperref}

\usepackage{tikz}
\definecolor{lime}{HTML}{A6CE39}
\DeclareRobustCommand{\orcidicon}{%
  \begin{tikzpicture}
  \draw[lime, fill=lime] (0,0) 
  circle [radius=0.16] 
  node[white] {{\fontfamily{qag}\selectfont \tiny ID}};
  \draw[white, fill=white] (-0.0625,0.095) 
  circle [radius=0.007];
  \end{tikzpicture}
  \hspace{-2mm}
}

\foreach \x in {A, ..., Z}{%
  \expandafter\xdef\csname orcid\x\endcsname{\noexpand\href{https://orcid.org/\csname orcidauthor\x\endcsname}{\noexpand\orcidicon}}
}

\usepackage{txfonts}

\begin{document}

   \title{Signatures and bias assessment of rotation in galaxy cluster members}

   \titlerunning{Rotation in galaxy clusters}
   \authorrunning{Castellani, Ferrami et al.}
    \author{D. Castellani \inst{1\thanks{\email{davide.castellani1@studenti.unimi.it} }} \orcidD{}
    \and
    G. Ferrami\inst{2\thanks{\email{gferrami@student.unimelb.edu.au} }} \orcidA{}
    \and
    C. Grillo \inst{1}\fnmsep\inst{3} \orcidC{}
    \and
    G. Bertin \inst{1}  \orcidB{}
    }

    \institute
    {
    Dipartimento di Fisica, Università degli Studi di Milano, via Celoria 16, I-20133 Milano, Italy
    \and
    School of Physics, University of Melbourne, Parkville, VIC 3010, Australia
    \and
    INAF -- IASF Milano, via A. Corti 12, I-20133 Milano, Italy
    }
    
   \date{Received -; accepted -}

\abstract 
   {In dynamically relaxed galaxy clusters, the galactic component is typically assumed to have zero or negligible mean motions.}
   {We investigate the possible presence of systematic rotation in the member galaxies of a sample of 17 nearby ($z<0.1$), rich (at least 80 identified members) Abell clusters. 
   We also assess the extent to which low-number statistics may influence the recovery of the rotation parameters.}
   {Following the methods often used in the context of globular clusters and of clusters of galaxies, we estimate a representative value of the systematic rotation velocity ($v_\mathrm{rot}$) and the position angle (PA) of the projected rotation axis for the set of spectroscopically confirmed member galaxies within 1.5 Mpc from the centre of each cluster.
   We study the robustness of our rotational velocity measurements as a function of the number of galaxies $N$ included in the analysis with a bootstrapping technique.}
   {Eight clusters with sufficiently abundant and regular data (A1367, A1650, A2029, A2065, A2142, A2199, A2255 and A2670) exhibit a significantly high rotational velocity, when compared to their velocity dispersion ($v_\mathrm{rot}/\sigma\geq 0.15$). Interestingly, three of them (A1650, A2029 and A2199) {are confirmed to be cool-core, relaxed clusters with no evidence of recent mergers, as suggested by X-ray observational data}. We also find a general tendency to overestimate the value of $v_\mathrm{rot}$ when the number of galaxies with measured velocities is reduced, for which we put forward an analytical justification. This bias mainly affects slowly rotating clusters: we find that clusters with $0.15 \leq v_\mathrm{rot}/\sigma \leq 0.20$ require at least 120 galaxies with measured velocities to limit the percentage error to less than $\sim10\%$, while for rotating clusters with $v_\mathrm{rot}/\sigma \approx 0.25$, $\sim 55$ kinematic data points are sufficient to achieve the same accuracy.}
   {}


   \maketitle

\section{Introduction}\label{sec:intro}

Galaxy clusters are the largest and most massive gravitationally bound objects observed in the Universe. Their large fraction of dark matter (DM) and highly ionised plasma (the intracluster medium, or ICM; often referred to as ``the gas'') makes them an ideal workbench for both observational cosmology \citep[e.g., see][]{ Reiprich2002, Allen2011} and X-ray astronomy \citep{Sarazin1988}. Even though they contribute by only a small percentage to the total cluster mass, galaxies remain a valuable tool for investigating the dynamical state of these systems.

Both the gas and the galactic components of most relaxed clusters are traditionally considered to be fully pressure-supported, in the sense that their kinetic energy is taken to be in the form of random motions of their constituents.
The ICM is frequently assumed to be in hydrostatic equilibrium, which provides a convenient model to derive the total mass of the cluster \citep[for a review, see][]{Ettori2013}. Similarly, galaxies in clusters are generally considered (as a collisionless component) to be in a quasi-stationary state of dynamical equilibrium with no or negligible systematic motions \citep[see, e.g.,][]{Carlberg1997}. In practice, such a picture largely derives from the lack of detailed spectroscopic studies, much like for the case of globular clusters non-rotating King models remained the standard paradigm until deeper kinematical studies became available \citep[see ][]{DjorgovskiMeylan1994, Zocchi2012, Bianchini2013}. It follows that a convincing detection of systematic rotation of galaxy clusters would change significantly our current picture of these systems, pushing forward the current understanding of their formation, evolution, and the role of the interaction with their surroundings. Cluster rotation could have different origins, such as tidal torques \citep{Peebles1969} or off-axis mergers \citep{Ricker1998, RoettigerFlores2000}. In other words, the existence of non-thermal dynamics in these systems might call for a significant revision of the standard cosmological framework, as a definitive understanding of the presence of angular momentum in the Large Scale Structure of the Universe still remains elusive. 

There is a growing effort to quantify and characterise the various non-thermal pressure contributions of the ICM, such as turbulent and bulk motions, including rotation \citep{Bianconi2013, Gianfagna2021, Ettori2022, Bartalesi2024, Bartalesi2025}. 
Correspondingly, a limited number of studies have hinted at the possible existence of non-negligible systematic rotation of the galactic component \citep{HwangLee2007, Tovmassian2015}, and mostly within the central regions of some specific clusters \citep{MaterneHopp1983, Ferrami2023}. An interesting example is Abell 2107, for which there is some evidence of rotation for both the ICM \citep{LiuTozzi2019} and the galaxies \citep{OegerleHill1992, Kalinkov2005, Song2018}.

Furthermore, most of the past investigations that looked for rotation of the galactic component have relied (in the best cases) on only a few tens of data points with measured kinematics and have primarily focused on the asymmetries of the velocity distribution in the relevant position-velocity diagrams \citep{Smith1936, Rood1972, Dressler1981}. However, in recent years, the significant increase in the availability of high-quality kinematic data for galaxies in clusters has encouraged the use of more accurate and robust rotation-detecting methods \citep[e.g., see][]{HwangLee2007, Ferrami2023}. This approach has already been proven to be successful in the parallel field of globular clusters, where a similar improvement in the quantity and quality of the data made it clear that many of these structures are partially rotation-supported \citep[see, e.g.,][]{Bianchini2013, Kacharov2014, Leanza2022}. 

As a collisionless component, it is believed that galaxies in clusters have reached a state of partial relaxation via the mechanism of violent relaxation \citep[see][]{Lynden-Bell1967}. While this process is not necessarily associated with the production of rotational motion, it leaves the core regions of the cluster generally more relaxed than the outer regions. Combined with the new developments on galaxy and globular clusters mentioned above, this suggests that when present, systematic rotation may be expected to be differential, with the central region typically rotating rigidly, while the outskirts may not rotate at all.

This paper has two main objectives. First, we look for evidence of systematic rotation in the galactic components of a sample of Abell clusters; this involves the use of a well-known algorithm that evaluates the rotational velocity and the projected position angle of the rotation axis. Second, we aim at clarifying and quantifying the "low-number bias" that we mentioned above and assess its impact on the estimates of rotational support as a function of the number of kinematic data points considered.

This paper is organised as follows. The sample of clusters with the relevant photometric and spectroscopic data is presented in Section \ref{DataSection}. In Section \ref{MethodsSection}, we outline the methods of investigation used for both the search of systematic rotation and the evaluation of the related low-number bias. Then, in Section \ref{ResultsSection}, we show the results of the analysis followed by their discussion in Section \ref{DiscussionSection}. The final conclusions and remarks are reviewed in Section \ref{ConclusionsSection}.

\section{Data samples}\label{DataSection}
In this paper, the analysis and comparison of our results with those of previous studies is carried out on two different datasets, collected with different instruments.

\subsection{Sample for rotation detection}
The search of rotation signatures is mainly conducted with the sample of clusters presented in \citet{DEugenio2015}. They discarded the clusters at redshift $z>0.1$ and those with richness $\mathcal{R}<2$ from the \citet{Abell1989} catalogue, resulting in a sample of 20 clusters. From \citet{DEugenio2015}, we have extracted the following properties, shown in Table \ref{SampleTable}.
\begin{itemize}
    \item The coordinates of the cluster centre, which hereafter we assume to be those of their BCGs. When the second brightest galaxy falls within 0.5 mag from the brightest one, the cluster centre is taken to be the mid-point between the two.
    \item The estimated value of the cluster redshift $z$. 
    \item The angular distance from the cluster centre $R$ (in arcmin) that corresponds to 1.5 Mpc at the cluster redshift, calculated with the Planck cosmology \citep[$H_0=67$ km s$^{-1}$ Mpc$^{-1}$, $\Omega_m=0.32$, $\Omega_\Lambda=0.68$, see][]{PlanckCollabo2020}. This radius is employed as a standard scale within which we look for rotation. {We opted for this particular value because, for typical clusters, it manages to balance the two effects mentioned at the end of Sect.\;\ref{sec:intro}. A larger radius would likely dilute the rotation signal, whereas a smaller one would reduce the number of member galaxies, preventing a more statistically robust analysis.} Note that these radii are calculated under the assumption that the clusters do not have any significant peculiar motion with respect to the Hubble flow.
\end{itemize} 

\begin{table}
    \centering
    \caption{Basic properties of the two cluster samples.}
    \label{SampleTable}
    \begin{tabular}{lcccc}
    \hline\hline
    ID    & R.A.        & Dec.     & $z$      & $R$ \\
        & (J2000)   & (J2000)  &    & (arcmin)\\
    \hline
    A16         & 00:16:46.30         & 06:44:39.84                   & $0.084$         & $16.09$                    \\
    A168        & 01:15:09.79          & 00:14:50.64                   & $0.045$         & $28.52$                    \\
    A1035       & 10:32:07.20         & 40:12:33.12                   & $0.080$        & $16.80$                    \\
    A1186       & 11:13:51.36        & 75:23:39.84                   & $0.079$        & $16.95$                    \\
    A1190       & 11:11:46.32        & 40:50:41.28                   & $0.079$        & $16.89$                    \\
    A1367       & 11:44:29.52        & 19:50:20.40                   & $0.021$        & $58.29$                    \\
    A1650       & 12:58:46.32        & $-0$1:45:10.80                 & $0.085$        & $15.97$                    \\
    A1656       & 12:59:48.72        & 27:58:50.52                   & $0.023$        & $54.23$                    \\
    A1775       & 13:41:55.68        & 26:21:53.28                   & $0.072$        & $18.38$                    \\
    A1795       & 13:49:00.48         & 26:35:06.72                   & $0.062$         & $21.14$                    \\
    A1904       & 14:22:07.92         & 48:33:22.32                   & $0.071$        & $18.76$                    \\
    A2029       & 15:10:58.80        & 05:45:42.12                   & $0.077$         & $17.43$                    \\
    A2065       & 15:22:42.72        & 27:43:21.36                   & $0.072$        & $18.40$                    \\
    A2142       & 15:58:16.08        & 27:13:28.56                   & $0.090$        & $15.10$                    \\
    A2151       & 16:05:14.88         & 17:44:54.60                   & $0.037$        & $34.41$                   \\
    A2199       & 16:28:36.96        & 39:31:27.48                   & $0.030$        & $41.80$                    \\
    A2244       & 17:02:43.92         & 34:02:48.48                   & $0.099$         & $13.82$                    \\
    A2255       & 17:12:30.96        & 64:05:33.36                   & $0.081$         & $16.61$                   \\
    A2256       & 17:03:43.44         & 78:43:02.63                   & $0.060$         & $21.82$                    \\
    A2670       & 23:54:10.08        & $-$10:24:18.00                & $0.076$       & $17.56$                    \\
    \hline
    AS1063      & 22:48:43.97        & $-$44:31:51.16                  & $0.348$       & $4.91$                   \\
    M1206       & 12:06:12.15        & $-$08:48:03.37                  & $0.435$         & $4.28$                   \\
    A370        & 02:39:52.93         & $-$01:34:37.00                  & $0.375$        & $4.68$                   \\
    \hline
    \end{tabular}
    \tablefoot{Coordinates of the cluster centres are given in the usual units of the equatorial coordinate system, i.e. hours:minutes:seconds for R.A. and degrees:arcmin:arcsec for Dec.}
\end{table}

The optical and spectroscopic data of this sample are taken from the SDSS DR18 \citep{Almeida2023}. For each cluster, we select the right ascension and declination of the galaxies within the angular radius $R$. We then extract only the most reliable spectroscopic redshifts (\textit{zWarning} = 0), listed with "GALAXY" spectroscopic type. Membership selection is made by performing a Gaussian fit to the redshift distribution in the whereabouts of the $z$ listed in Table \ref{SampleTable}. Note that fast rotating clusters might show a strongly bi-peaked redshift distribution, but no major deviations from the usual Gaussian distribution of redshifts are expected for slow rotators. This allows us to adopt the described fitting procedure even in the presence of a relatively low rotational velocity.
 
We then obtain a new estimate of the cluster redshift ($z_\mathrm{cl}$) as the mean of the Gaussian curve. Therefore, the fitted standard deviation ($\sigma_z$) can be converted into the global velocity dispersion of the galactic component with
\begin{equation}
    \sigma_\mathrm{fit} = \frac{c}{1+z_\mathrm{cl}}\,\sigma_z\:.
\end{equation}
Finally, we select member galaxies by means of a $3\sigma_\mathrm{fit}$-clipping of the line-of-sight velocity relative to the cluster mean, i.e., a galaxy with measured redshift $z_\mathrm{g}$ is considered a member if its line-of-sight velocity $v$ is:
\begin{equation}
    |v|=|z_\mathrm{g}-z_\mathrm{cl}|\,c<3\sigma_\mathrm{fit}\:.
\end{equation}
For the sample for rotation detection, the fitted cluster redshift, the final total number of member galaxies and the cluster global velocity dispersion are shown in the second, third and fourth columns of Table \ref{tab:rot_fit}.
Note that the sample for rotation detection is reduced to 17 clusters because SDSS does not have spectroscopic data for A16, A1186 and A2256.

\begin{table}
    \centering
    \caption{Fitted rotation parameters.}
    \label{tab:rot_fit}
    \begin{tabular}{lcccccc}
    \hline\hline
    ID  & $z_\mathrm{cl}$  & $N_\mathrm{tot}$    & $\sigma_\mathrm{fit}$     & $v_\mathrm{rot}$      & PA     &     $v_\mathrm{rot}/\sigma_\mathrm{fit}$ \\
    & & & (km/s) & (km/s) & ($\degr$) & \\
    \hline
    A16          &  \ldots  &     \ldots     &    \ldots     &    \ldots     &    \ldots  &   \ldots          \\
    A168         &  0.045   &     155     &    542      &    15     &    253    &   0.028           \\
    A1035        &  0.073  &     94      &    1981      &     834    &    174    &   0.421           \\
    A1186        &  \ldots   &     \ldots     &    \ldots     &    \ldots     &    \ldots  &   \ldots   \\
    A1190        &  0.075   &     69     &     754     &    64     &   80     &     0.085         \\
    A1367        &  0.021  &     209      &    802      &    117     &    147     &    0.146          \\
    A1650        &  0.084  &     55      &    763      &    176     &    109     &     0.231         \\
    A1656        &  0.023   &     500     &    1039      &    48     &    224    &     0.046         \\
    A1775        &  0.075 &     66       &     591     &    46     &    260    &      0.077        \\
    A1795        &  0.063  &     87      &     883     &    48     &    288    &     0.054         \\
    A1904        &  0.072  &     89      &     993     &    92     &    177    &     0.093         \\
    A2029        &  0.078  &     105      &     1282     &    229     &    58    &    0.179          \\
    A2065        &  0.072   &     126     &     1567     &  257     &   89     &    0.165          \\
    A2142        &  0.090  &     129      &    940      &   181      &    165    &     0.193         \\
    A2151        &  0.036  &     211      &    809      &    214     &    235    &     0.265         \\
    A2199        &  0.030   &     222     &    742      &     154    &    78    &     0.208         \\
    A2244        &  0.099   &     75     &     1233     &    140     &   183     &    0.114          \\
    A2255        &  0.080   &     74     &     1162     &    291     &    6     &     0.250         \\
    A2256        &  \ldots   &     \ldots     &    \ldots     &    \ldots     &    \ldots  &   \ldots   \\
    A2670        &  0.076   &     98     &    924      &     190    &   253     &      0.206        \\
    \hline
    \end{tabular}
    \tablefoot{$z_\mathrm{cl}$ is the estimate of the cluster redshift, determined as the mean of the Gaussian fit to the redshift distribution. In the third column, $N_\mathrm{tot}$ refers to the number of member galaxies within the scale radius $R$ (see Table \ref{SampleTable}). The fourth column shows the global velocity dispersion, estimated from the Gaussian fit to the redshift distribution. The fifth and sixth columns display, respectively, the rotation amplitude and the projected position angle of the axis determined in the way described in Section \ref{MethodsSection}. For the three clusters A16, A1186 and A2256, SDSS does not provide spectroscopic data; therefore, the sample for rotation detection is reduced to 17 clusters.}
\end{table}

\subsection{Bias-test sample}\label{bias-test_sample_Section}
Additional tests of the low-number bias have been performed on a different dataset, consisting of three rich and more distant ($z\sim 0.4$) clusters (Abell S1063, MACS J1206.2$-$0847 and Abell 370) with high-quality data, for which previous studies have suggested the possibility of rotation and identified them as good candidates for this type of study. 

The dynamics of Abell S1063 ($z\approx0.35$) and MACS J1206.2$-$0847 (hereafter MACS J1206; $z\approx 0.44$) have been studied extensively by \citet{Ferrami2023}.
The data primarily come from the CLASH-VLT program \citep{Rosati2014}. It builds on the Cluster Lensing And Supernova survey with Hubble \citep{Postman2012}, a multi-wavelength photometric survey that observed 25 massive galaxy clusters and on a spectroscopic campaign using the VIMOS instrument \citep{LeFevre2003} on the Very Large Telescope. This campaign was further enhanced by observations from the MUSE \citep[Multi-Unit Spectroscopic Explorer,][]{Bacon2010} spectrograph, providing comprehensive data on the kinematics of galaxies within the innermost regions of the clusters \citep[$\sim 2$ arcmin$^2$ for Abell S1063 and $\sim 2.6$ arcmin$^2$ for MACS J1206; see, respectively,][]{Caminha2016,Caminha2017}. After membership selection, the final dataset consists of 1215 galaxies for Abell S1063 and 658 galaxies for MACS J1206, each covering an area of $25\times25$ arcmin$^2$ (or, equivalently, of $7.7\times7.7$ Mpc$^2$ and $8.9\times8.9$ Mpc$^2$, respectively).

For Abell 370 ($z\approx 0.38$), we use the data sample assembled by \citet{Lagattuta2022}, who studied the structure and kinematics of this strong-lensing cluster. This catalogue consists of 416 galaxy redshifts, spanning both the central region and the outskirts, targeted with the MUSE spectrograph on multiple visits. The initial spectroscopic dataset is based on a narrower field, covering a total area of 4 arcmin$^2$ around the identified centre \citep{Lagattuta2019}.
\citet{Lagattuta2022} included shallower MUSE observations covering more external regions, resulting in a final central coverage of an area of 14 arcmin$^2$ (or $1.46$ Mpc$^2$) around the core. The photometric data of Abell 370 is composed of high-resolution images mainly from the Hubble Frontier Fields programme \citep[HFF,][]{Lotz2017} for the cluster core, integrated with less deep measurements on the outskirts from the Beyond Ultra-deep Frontier Fields And Legacy Observations \citep[BUFFALO,][]{Steinhardt2020} project, with coverage of nearly 6 × 6 arcmin$^2$ ($1.9\times 1.9$ Mpc$^2$).

The basic properties of this second cluster sample are recorded in the bottom part of Table \ref{SampleTable}.

\section{Methods}\label{MethodsSection}

\subsection{Rotation measurements}\label{RotationDetectionMethodSubsection}
\begin{figure*}
    \centering
  \includegraphics[width=\linewidth, keepaspectratio]{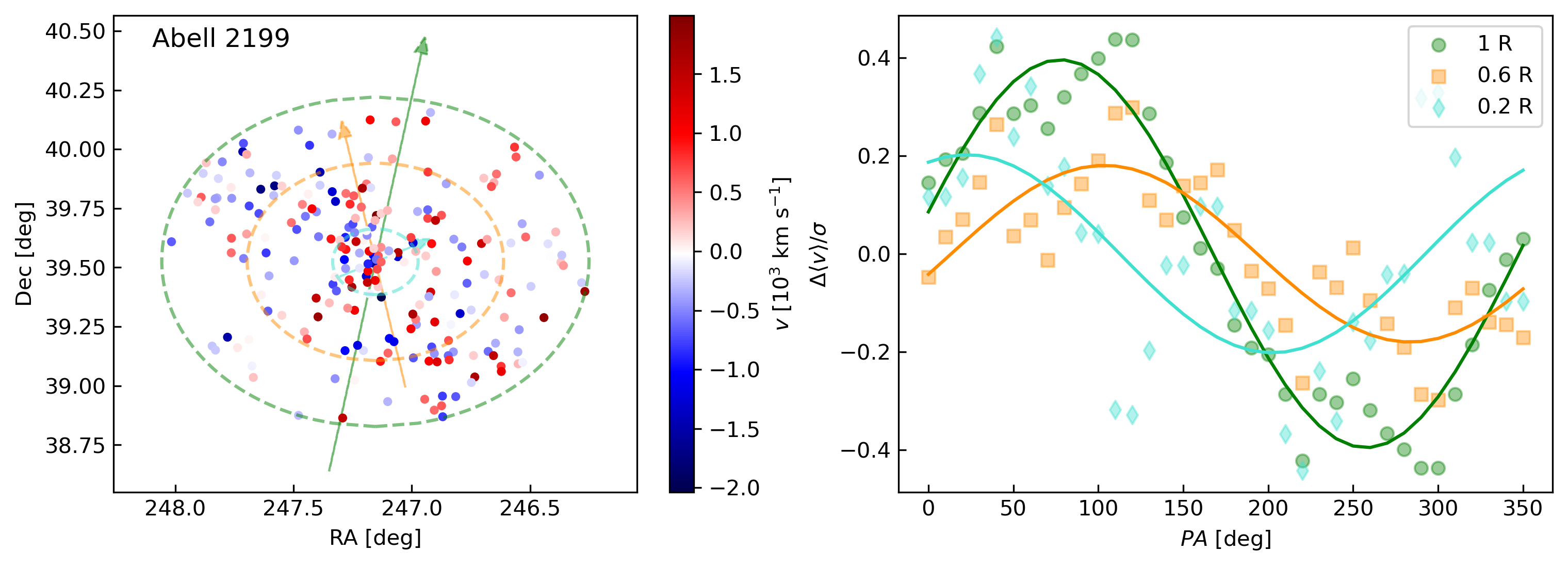}
  \caption{Example of the fitting procedure used to search for systematic rotation of the galactic component in clusters of galaxies. Left: projected spatial distribution of galaxies of Abell 2199, colour-coded according to their line-of-sight velocity relative to the cluster mean ($v$). The dashed contours represent three different cuts within which we performed the test, in order to monitor whether the velocity amplitude would increase/decrease and how the axis position angle changes with radius. Right: the corresponding sine fits, from which we extracted the parameters related to the largest scale ($R$) reported in Table \ref{tab:rot_fit}.}\label{fig:AVD_fitting}
\end{figure*}

The method we used to estimate the position angle of the rotation axis and the the rotational velocity has been employed in several previous studies, primarily on the available stellar kinematical data of globular clusters \citep[see, e.g.,][]{Cote1995, Bianchini2013, Kacharov2014, Leanza2022, Leanza2023}. It operates as follows.
The initial step is to subtract the mean velocity of the cluster from the line-of-sight velocity of the member galaxies. Then, the projected image of the cluster is divided into two halves by a line passing through the centre at a given position angle $\theta_i$. For each of the two resulting subsamples, the average line-of-sight velocity $\langle v\rangle_i$ is calculated and the difference between them is then plotted against $\theta_i$. Subsequently, the position angle is increased by $10\degr$ (so that $\theta_{i+1} = \theta_{i}+10\degr$), and the same procedure is repeated until the entire $360\degr$ are covered. Finally, the resulting pattern is fitted with a sine function so that the angle at which the difference between the mean velocities reaches its maximum corresponds to the position angle of the rotation axis, while the amplitude of the sine provides an estimate of a representative value of the systematic rotation velocity. An example of this procedure is shown in Fig. \ref{fig:AVD_fitting}. (In passing we note that, in the left frame of the figure, the contours are not circular, because of the coordinates adopted in the map).

This method has been applied to the galactic component of Abell S1063 and MACS J1206 by \citet{Ferrami2023}. Similar methods are described and used in the literature; in particular, they have been used to study the possible presence of systematic rotation in clusters of galaxies \cite[see, e.g.,][]{Dressler1981, HwangLee2007}. We decided to test the methods used in \citet{HwangLee2007} and \citet{Ferrami2023} on a set of simplified simulated galaxy clusters, for which the rotational velocity $v_\mathrm{rot}$ and velocity dispersion $\sigma$ are known in advance, to determine which method is statistically more accurate in retrieving the correct rotation parameters as a function of $v_\mathrm{rot}/\sigma$. The specifics of the simulation and the results are presented in Appendix \ref{app:methods_testing}. It turns out that the method presented above is consistently better at recovering the projected orientation of the rotation axis, and it is probably preferable when considering clusters with relatively low rotational velocity ($v_\mathrm{rot}/\sigma\la 0.30$).

\subsection{Low-number bias}\label{BiasModelSubsection}
In carrying out our investigations we found that, somewhat unexpectedly, systems with smaller numbers of kinematical data points often appeared to exhibit a stronger signal of rotation, and thus we became aware that we might have to face a bias problem. Note that, as outlined in Section \ref{sec:intro}, there are two main reasons why some systems may have limited kinematic measurements: either due to insufficient spectroscopic observations, or because the analysis is restricted to the cluster core. Here, new observations might not be feasible simply because of lack of suitable targets in the region considered.
To investigate the extent to which the number of member galaxies influences the fitted rotation parameters, we repeatedly apply the rotation detection algorithm described above to randomly selected subsets of $N$ member galaxies of the sample of 17 clusters listed in the upper part of Table \ref{SampleTable}. Starting from the total sample of $N_\mathrm{tot}$ galaxies, {we progressively reduce the sample size by randomly removing 10 galaxies at a time, and extract $v_\mathrm{rot}$ 5000 times for each value of $N$.} Under the assumption that the value fitted when $N=N_\mathrm{tot}$ is consistent with the one calculated with infinite statistics (that is to say, the true rotation velocity), we study the trend of the average rotation velocity to velocity dispersion ratio as a function of $N$.

A simple model has been found to provide an approximate estimate of the bias introduced by the low-number statistics; for a derivation of the following Eq.\;(\ref{eq:ratio_as_N}) and for the definition of the correlation coefficient, the reader is referred to Appendix \ref{app:bias_derivation}.
If a cluster is rotating, the difference between the mean velocities calculated on each side of the rotation axis projected on the sky will be non-zero. Let us call $\Delta v$ half this mean velocity difference, as measured in a sample of $N$ galaxies.
For $N \rightarrow \infty$ this value approaches $v_\mathrm{rot}$. 
Similarly, we can define $\sigma_v$ as the velocity dispersion calculated for a sample of $N$ galaxies, and $\sigma$ its limit for large $N$.
Then, the ratio $\langle|\Delta v|\rangle/\sigma_v$ as a function of $N$ is

\begin{equation}\label{eq:ratio_as_N}
\frac{\langle|\Delta v|\rangle}{\sigma_v} = 
\left[
\frac{v_\mathrm{rot}}{\sigma}
\erf{\left(\frac{v_\mathrm{rot}}{\sigma}\sqrt{\frac{N}{\hat\rho}}\right)} +
\sqrt{\frac{\hat\rho}{\pi N}}
\exp\left(-\frac{v_\mathrm{rot}^2}{\sigma^2}\frac{N}{\hat\rho}\right)
\right] g(N) \:,
\end{equation}
where $\hat\rho = 1/(1+\rho)$ depends on the correlation coefficient $\rho$ between the samples on each side of the rotation axis, and
\begin{equation}\label{eq:g_N}
g(N)= \sqrt{\frac{N-1}{2}}
\frac{\Gamma((N-1)/2)}{\Gamma(N/2)} \:.
\end{equation}
Equation (\ref{eq:ratio_as_N}) tells us that $\langle|\Delta v|\rangle/\sigma_v$ as a function of $N$ depends on two parameters: the limit ratio ${v_\mathrm{rot}}/{\sigma}$ and the correlation coefficient $\rho$. As previously mentioned, the first parameter is specific to each cluster and is approximated using the value obtained when all the galaxies were considered. In the case of the second parameter, under the assumption that the correlation coefficient is the same for each cluster, we calibrate its value in the following section using the eight clusters where significant evidence for systematic rotation has been found.

\section{Results}\label{ResultsSection}
\subsection{Rotation detection and model calibration}\label{Rotation detection and model calibration Subsection}

\begin{figure}
    \centering
  \includegraphics[width=\linewidth, keepaspectratio]{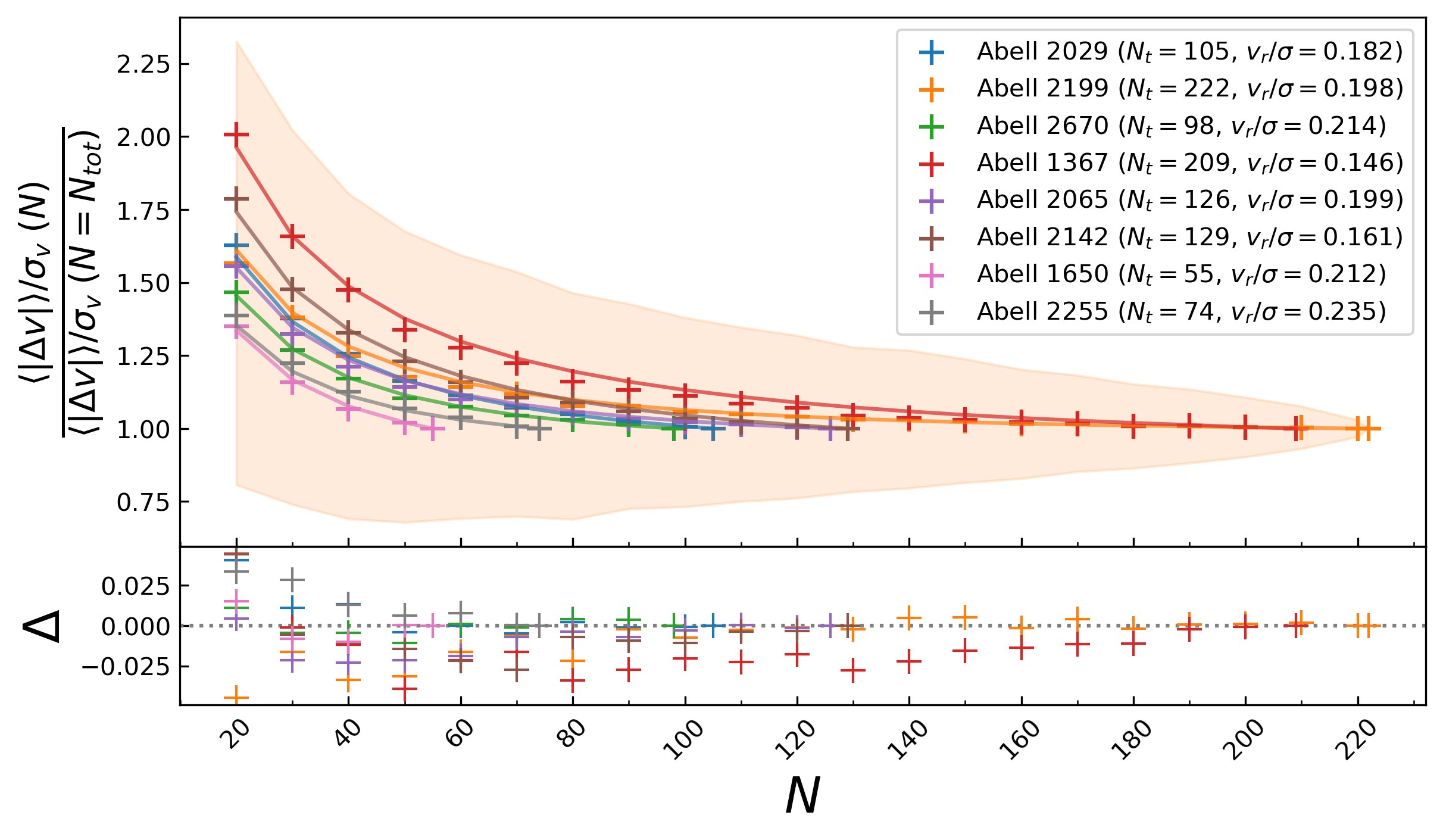}
  \caption{Upper panel: the low-number bias for the eight clusters selected in this paper compared to the related analytical curve with the best-fit $\rho=-0.788$. {To illustrate the statistical uncertainty without overcrowding the graph, we only show the A2199 one-sigma error band for reference.} Lower panel: the residuals $\Delta$ between the data points and the analytical curves. In the legend, we use $N_t=N_\mathrm{tot}$, $v_r=v_\mathrm{rot}$ and $\sigma$ calculated as the standard deviation of the full set of galaxies (this is what ultimately causes the difference between the values of $v_r/\sigma$ reported here and the $v_\mathrm{rot}/\sigma_\mathrm{fit}$ in Tab.\;\ref{tab:rot_fit}, where $\sigma_\mathrm{fit}$ is extracted from the Gaussian fit of the redshift distribution).
  The case of $N=10$ was excluded from the analysis, as undesired additional systematic effects are expected to appear with such a small number of galaxies.}\label{fig:Bias_remaining}
\end{figure}

As noted at the beginning of Sect. \ref{RotationDetectionMethodSubsection}, Fig. \ref{fig:AVD_fitting} illustrates the method that we have applied to derive the rotation parameters listed in Table \ref{tab:rot_fit}. The same figure also clarifies what happens in the application of the method when, for a given cluster with a given set of data, different cuts are performed in order to consider smaller and smaller regions. In the case shown in Fig. \ref{fig:AVD_fitting}, the cuts have been made at $1 R$, $0.6 R$, and $0.2 R$. In the absence of the low-number bias, this procedure might provide indications on the radial dependence of rotation. The low-number bias tends to confuse the picture, because the different cuts define datasets with different values of $N$ (see comments at the beginning of Sect. \ref{BiasModelSubsection}). In this respect, one should note that the present method leads to an estimate of the typical rotation speed, as a {\it linear velocity}; for rigid rotation, that is constant {\it angular velocity}, the typical rotation speed decreases linearly with decreasing radius. The combination of these factors does not allow for an unambiguous interpretation of the trend shown by the three curves on the right panel of Fig. \ref{fig:AVD_fitting}.

In 10 of the 17 clusters in the sample for rotation detection, the fitted parameter values indicate a significant rotation support over random motions ($v_\mathrm{rot}/\sigma\ga 0.15$, see Table \ref{tab:rot_fit}). However, a high value of this parameter does not necessarily guarantee that the cluster is actually rotating. In fact, a similar signal may be caused by various factors, including the following:
\begin{enumerate}
    \item The cluster could consist of two (or more) separate clusters at slightly different angular positions in the sky. This is the case for systems exhibiting a distinct sub-clustering of galaxies, often associated with a strongly two-horned distribution of redshifts. 
    \item The cluster may be the result of a recent (or ongoing) merging event and may not have reached a quasi-equilibrium state yet. Whether or not off-axis mergers could mimic the presence of systematic rotation, the search for angular momentum in clusters of galaxies is best carried out in systems that appear to have already settled in a quasi-equilibrium state. In this respect, X-ray data of the intracluster medium offers a particularly useful probe of the dynamical state of clusters, with unrelaxed systems displaying evidence of disturbances such as shock fronts, filaments, or irregular isophotes.
    \item The dataset could contain an insufficient number of data points with measured kinematics, and therefore suffer from the low-number bias mentioned in Sect.\;\ref{BiasModelSubsection}. 
\end{enumerate}

In our subsample of 10 clusters with an indication of rotation, for the following analysis we discard those that previous investigations have shown to consist of close but distinct substructures: this is the case of A1035 \citep[see][]{KopylovKopylova2007} and A2151 \citep[see][]{Monteiro-Oliveira2022}.
This leaves us with eight clusters with a fairly relaxed optical morphology and whose redshift distribution exhibits indications of systematic rotation (A1367, A1650, A2029, A2065, A2142, A2199, A2255 and A2670). Interestingly, \citet{Bartalesi2025} found that the current X-ray data leave room for significant systematic rotation of the ICM in three clusters of their dataset, the most promising candidate being A2255. Furthermore, three of the eight clusters selected above are considered to be typical, cool-core, relaxed clusters with regular X-ray morphology and no evidence of recent merging events: A1650 \citep{Govoni2009}, A2029 \citep[see, e.g,][]{BuoteTsai1996} and A2199 \citep[see, e.g.,][]{Markevitch1999}. 

For the second objective of this paper - that is, to quantify the low-number bias - we will consider the eight clusters with a sufficiently regular optical morphology and significant evidence of systematic rotation.

Figure \ref{fig:Bias_remaining} shows the resulting trends of the average $\langle|\Delta v|\rangle/\sigma_v$ when gradually removing galaxies from the eight selected clusters ($5000$ resamplings for each $N$). The rotation parameter, normalised to its limit value when using all the galaxies, shows a steep increase as the number of kinematic data points is reduced.
We also plot the related analytical curves (Eq. \ref{eq:ratio_as_N}) and the corresponding residuals $\Delta$. The optimal value of $\rho$ has been determined by minimising the sum of the clusters mean square errors. For this optimisation process, the case with $N=10$ has been excluded, as the hypotheses at the basis of the simple model described in Appendix \ref{app:bias_derivation} may not be applicable in this context (for example, the Gaussianity of the distributions and the equal number of the galaxies on the two sides of the axis, see Appendix \ref{app:bias_derivation}). The final best-fit value of the correlation coefficient turned out to be $\rho_{\mathrm{opt}}=-0.788$. A more general approach is taken in Appendix \ref{app:simult_fit}, where both the limit ratio $v_\mathrm{rot}/\sigma$ and the parameter $\rho$ are fitted simultaneously. This different procedure confirms that the latter
coefficient remains approximately the same for each cluster. This justifies our original assumption that all clusters exhibit a similar correlation coefficient when considering the two subsamples of galaxies.

\begin{figure}
    \centering
  \includegraphics[width=\linewidth, keepaspectratio]{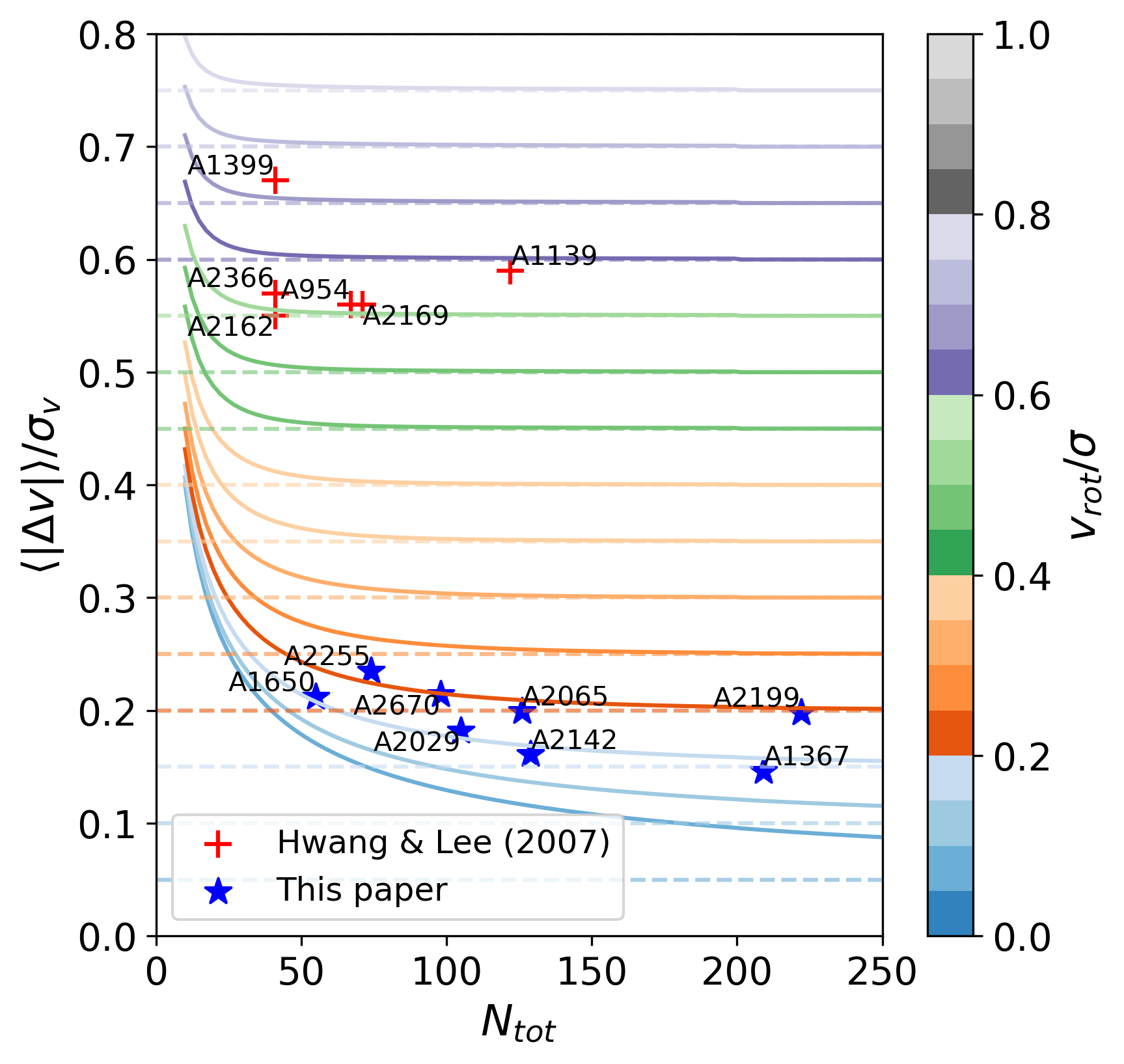}
  \caption{The different impact of the low-number bias between fast and slow rotating clusters. Because of their stricter selection criteria, \citet{HwangLee2007} found clusters whose rotational support is so high that they can be almost assumed to be bias-free, even though most of them have only a few tens of data points with measured kinematics.}\label{fig:Bias}
\end{figure}

As suggested by Fig. \ref{fig:Bias_remaining}, there appears to be an inverse relation between the magnitude of the rotation signal and the level of the bias, in the sense that the bias is low when the rotation signal is high. This is clarified further by Fig. \ref{fig:Bias}, where the analytical curves as a function of $N$ are plotted for different fixed true ratios $v_\mathrm{rot} / \sigma$ in the limit $N\to\infty$.
As the value of $v_\mathrm{rot} / \sigma$ increases, the influence of $N$ on the measured ratio diminishes. This is due to the fact that the folded Gaussian distribution of velocities undergoes only minimal alterations at its tails (see Appendix \ref{app:bias_derivation}). Furthermore, the values of the eight clusters employed in this study, along with those identified by \citet{HwangLee2007}, are superimposed on the curves. These last clusters exhibit minimal bias, as evidenced by their high rotational velocities.

\subsection{Bias testing with radial cuts}\label{sect:Bias testing with radial cuts}
Finally, we try to determine whether the bias affects the measurements in a similar way when the galaxies are removed from the projected distribution with progressively smaller radial cuts, rather than being randomly selected across the entire extent of the cluster. We make this test by using the three massive clusters with a large number of identified cluster members (Abell S1063, MACS J1206, and Abell 370) of the bias-test sample, introduced in Sect.\;\ref{bias-test_sample_Section}.
As noted earlier in Subsection \ref{Rotation detection and model calibration Subsection}, in relation to the interpretation of Fig.\;\ref{fig:AVD_fitting}, an increase in $|\Delta v|/\sigma_v$ when focussing on the central regions could be explained by a real rotation signal rather than the low-number bias, even though for rigid rotation the representative value of rotation decreases when the radius of the disc is decreased.

Another important issue that needs to be addressed is the fact that removing galaxies with consecutive radial cuts produces only one realisation of the cluster for each $N$, whereas in Fig.\;\ref{fig:Bias_remaining} each point is the average of 5000 resamplings. Therefore, the specific distribution of galaxies in each cluster introduces a large fluctuation in the $|\Delta v|/\sigma_v$ trend. In order to mitigate this effect, we adopted the following procedure.
\begin{enumerate}
    \item To be consistent with the previous tests, we focus our attention on $1.5$ Mpc as the outermost scale ($R$, see Table \ref{SampleTable}), containing $N_\mathrm{tot}$ galaxies.
    \item For every fixed value of $N$ increasing in steps of 10 up to $N_\mathrm{tot}$, we calculate the cut-off radius containing the $N$ galaxies.
    \item For each $N$, corresponding to a radial cut, we make 500 different resamplings by randomly removing 40\% of the galaxies. After trying several removal percentages, we settled, somewhat arbitrarily, on the value of 40\%, as this represents a good compromise between removing too few galaxies, which would not mitigate the $|\Delta v|/\sigma_v$ fluctuation, and removing too many galaxies, which would significantly alter the overall appearance of the cluster. We then calculate for each resampling its $|\Delta v|/ \sigma_v$ and take their average. Note that the resulting $\langle|\Delta v|\rangle/\sigma_v$ may be an overestimate, as we expect the low-number bias to come into play when using a subsample of galaxies.
    \item Finally, given this undesired effect introduced in the previous step, we correct for the low-number bias using the analytical model  with $\rho=-0.788$ calibrated in Sect.\;\ref{Rotation detection and model calibration Subsection}.
\end{enumerate}

 \begin{figure}
    \centering
  \includegraphics[width=\linewidth, keepaspectratio]{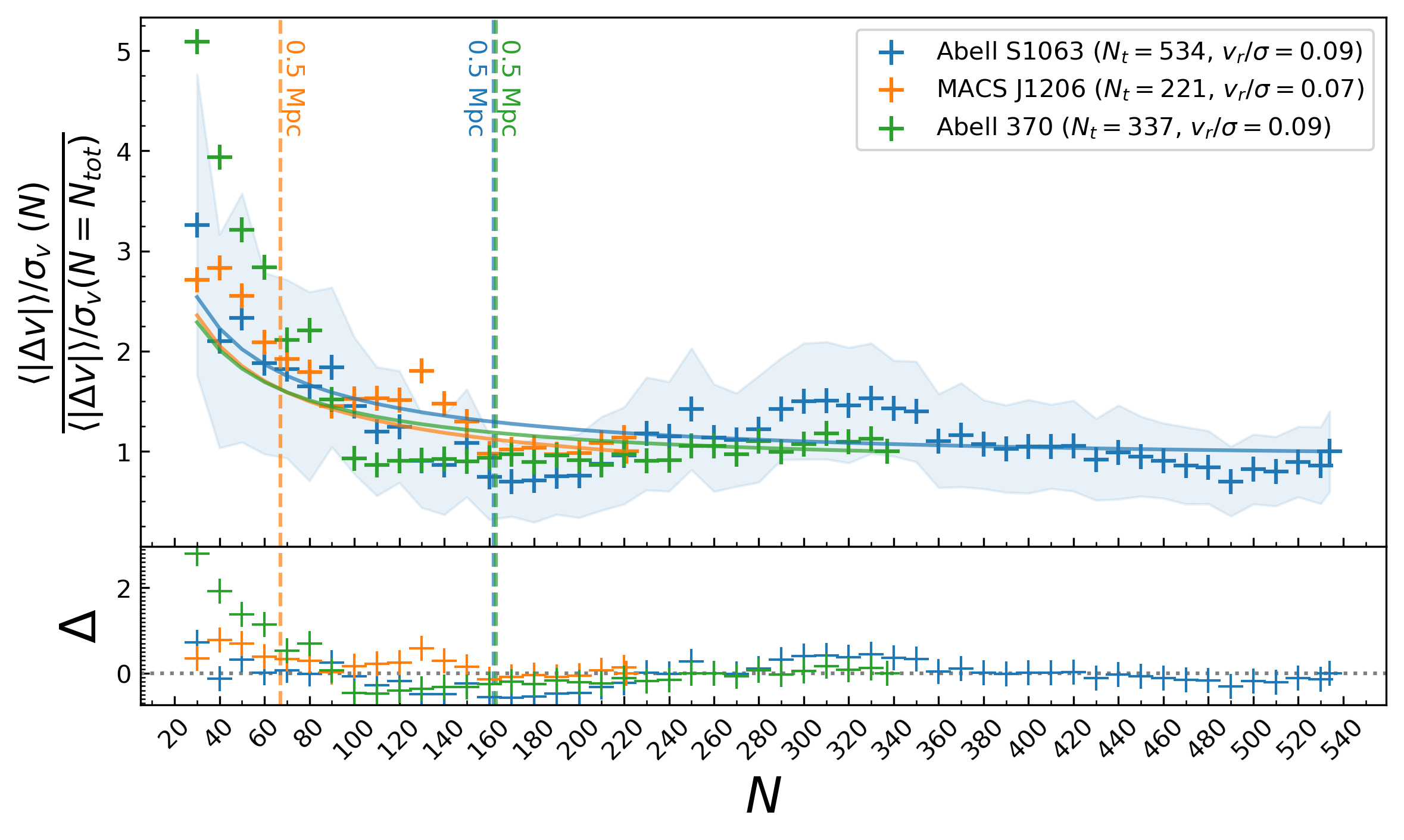}
  \caption{Rotation velocity over velocity dispersion trends when galaxies are removed by successive radial cuts in the three clusters with large number of identified members with measured line-of-sight velocities: Abell S1063, MACS J1206, Abell 370 (in the legend, we use $N_t=N_\mathrm{tot}$ and $v_r=v_\mathrm{rot}$). As in Fig.\;\ref{fig:Bias_remaining}, we display only the AS1063 one-sigma error band to illustrate resampling uncertainty without overcrowding the figure. While for the first two the low-number bias model reproduces the observed trend reasonably well, A370 deviates significantly in the central region, thus indicating convincing evidence for systematic rotation in its core. The vertical lines indicate the number of galaxies within a $0.5$ Mpc radius.}\label{fig:Bias_clusters}
\end{figure}

Figure \ref{fig:Bias_clusters} shows the final $\langle|\Delta v|\rangle/ \sigma_v$ trend for Abell S1063, MACS J1206 and Abell 370, after applying the procedure described above. In the first two cases, the measured ratios of rotation velocity over dispersion appear to roughly align with the trend predicted by our model of the bias. This supports the idea that, in these two clusters, if there is any systematic rotation of the member galaxies, it might be a negligible component of their dynamics.
At variance with these two clusters, Abell 370 shows a significant deviation from the predicted trend of the bias in the central regions; this apparently provides compelling evidence that the cluster core has a substantial amount of rotational support.

\section{Discussion}\label{DiscussionSection}
\begin{figure}
    \centering
  \includegraphics[width=\linewidth, keepaspectratio]{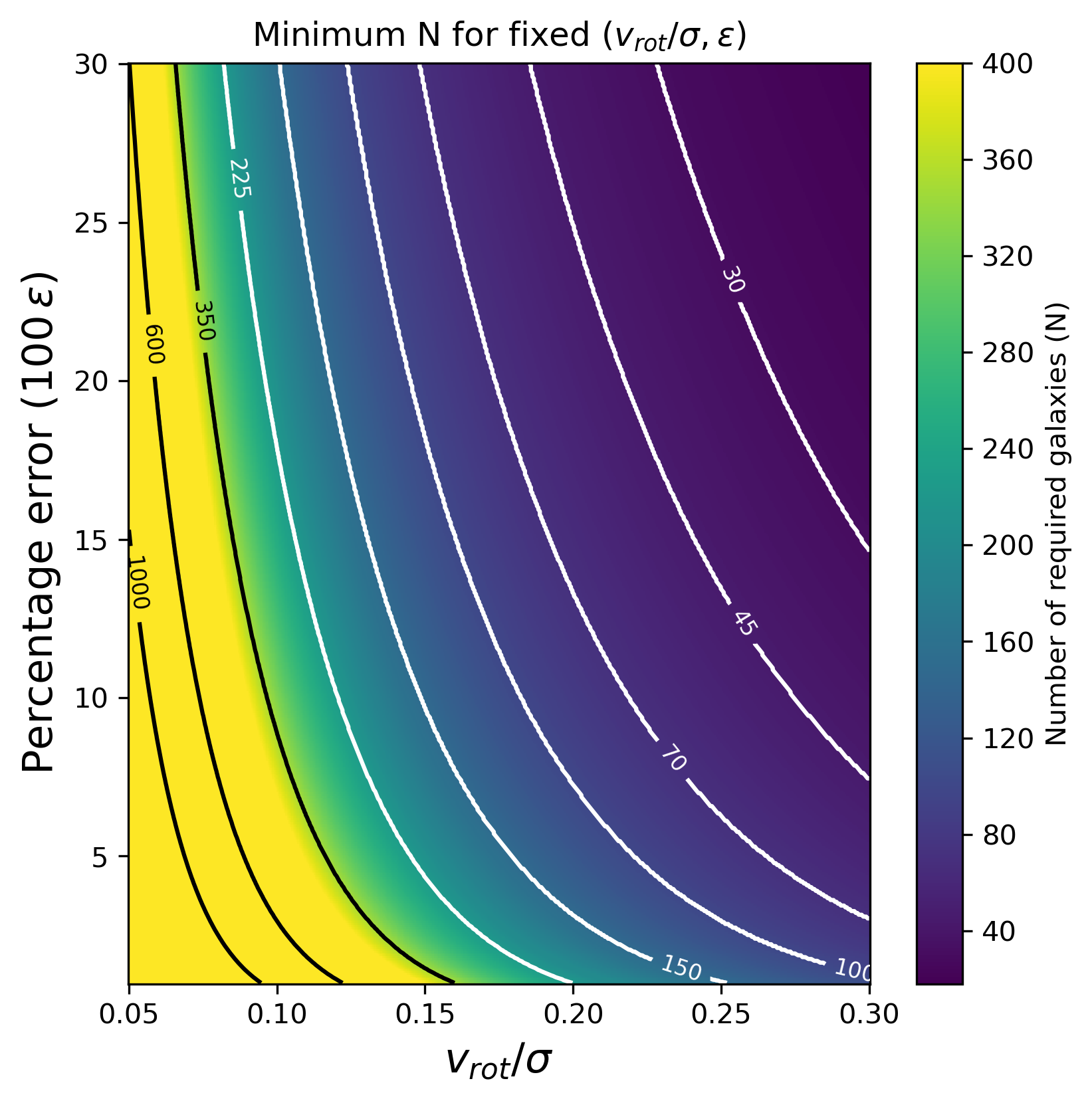}
  \caption{Minimum number of galaxies needed to achieve a relative error $\varepsilon$ in the case of a true ratio of $v_\mathrm{rot}/\sigma$.}\label{fig:minN}
\end{figure}

The extent to which the low-number bias can affect the correct recovery of the rotation parameters varies from case to case, as it depends mainly on the rotational velocity and the number of galaxies used to extract it. Using Eq.\;(\ref{eq:ratio_as_N}) and the definition of relative error
\begin{equation}\label{eq:increment}
    \varepsilon = \frac{\langle |\Delta v|\rangle / \sigma_v}{v_\mathrm{rot}/ \sigma} - 1\;,
\end{equation}
we can study this dependence by finding the minimum number of galaxies $N$ required to get an error on the order of $\varepsilon$ in the case of a true ratio $v_\mathrm{rot} / \sigma$.
The results are shown in Fig. \ref{fig:minN}. We find that galaxy clusters with $v_\mathrm{rot}/\sigma\ga 0.25$ are essentially bias-free when using around 100 galaxies or more; for the $0.15 \la v_\mathrm{rot}/\sigma\la 0.20$ range, it would be preferable to use more than 120 data points to limit the error to $\varepsilon\la 0.1$. Finally, it appears that slow-to-non rotators ($v_\mathrm{rot}/\sigma\la 0.15$) require a significantly larger number of galaxies (up to more than 300) to validate the rotation signal.

The low-number bias might be studied and should also play a role in the context of globular clusters, in those cases where the search for rotation is carried out with similar procedures. A real case example of how this bias can impact rotation measurements is studied in Appendix \ref{app:bias_globular_cluster}. However, for the study of star motions in globular clusters other sources of information are often available (in particular, proper motions for Galactic globular clusters).

\begin{figure}
    \centering
  \includegraphics[width=\linewidth, keepaspectratio]{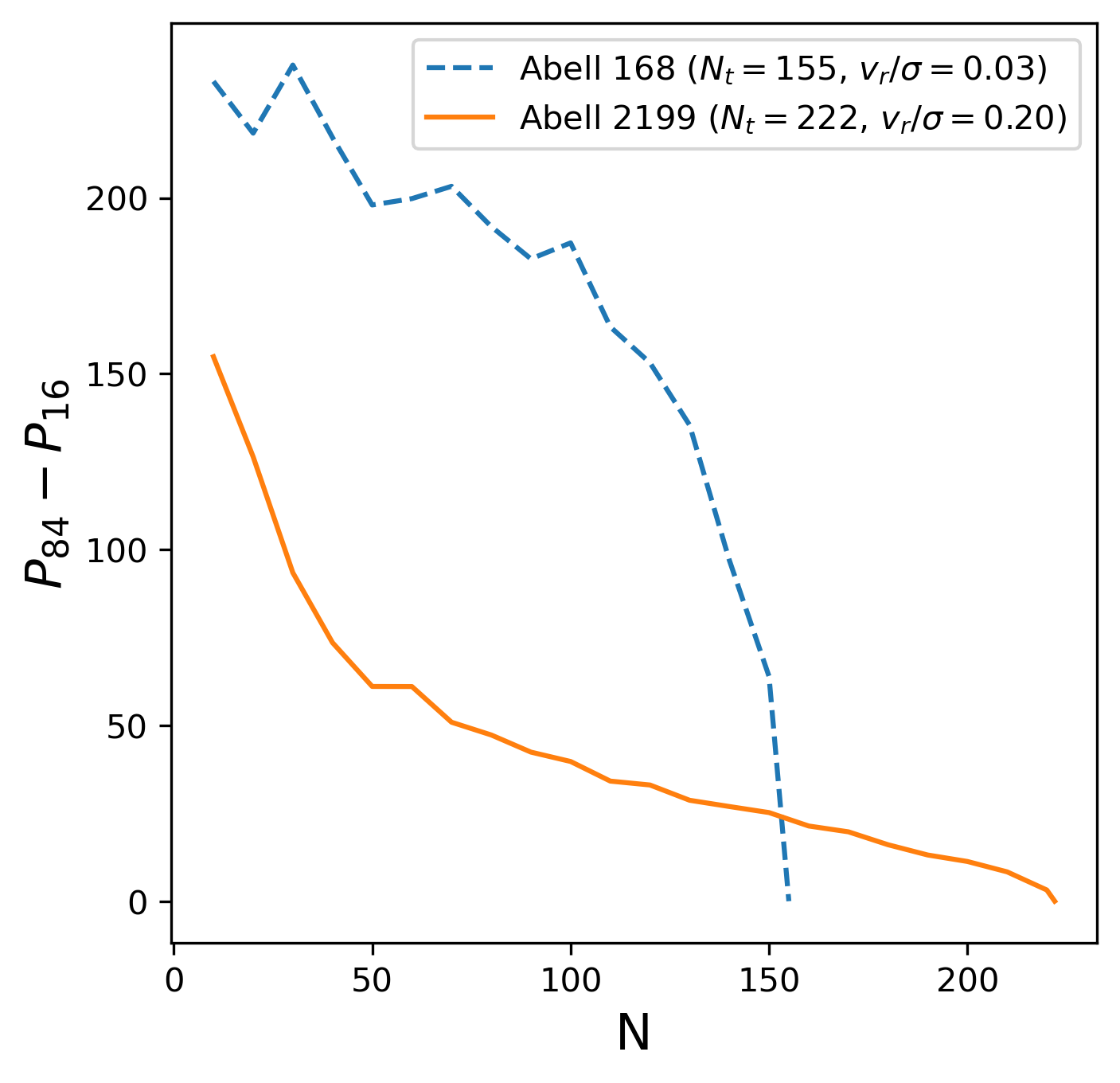}
  \caption{Difference between the 84$^\mathrm{th}$ and 16$^\mathrm{th}$ percentile of PAs for each $N$ of a non-rotating (Abell 168) and a rotating (Abell 2199) cluster (in the legend, we use $N_t=N_\mathrm{tot}$, $v_r=v_\mathrm{rot}$ and $\sigma$ calculated as the standard deviation of the full set of galaxies).}\label{fig:percentiles}
\end{figure}

A final point worth discussing is how the PA varies when we remove galaxies, either by bootstrapping or by radial cuts. As it can be expected, if there is indeed a systematic redshift distribution of the cluster member galaxies, the PA should remain fairly centred on a single value, even if we do not use the entire dataset. A narrower distribution of PAs is therefore a sign of a more stable and thus more reliable rotation measure. 
We have investigated this fact by keeping track of the difference between the 84$^\mathrm{th}$ and 16$^\mathrm{th}$ percentiles ($P_{84}-P_{16}$) of the distribution of PAs extracted from the random subsets of cluster galaxies used for the characterisation of the low-number bias (see Section \ref{BiasModelSubsection}, 5000 resamplings for each $N$). Representative results for a rotating and a non-rotating clusters are shown in Fig.\;\ref{fig:percentiles}. This shows that in both cases the confidence interval containing 68\% of the PAs becomes larger as fewer galaxies are used. However, the non-rotating cluster A168 shows a much steeper increase, suggesting the absence of a preferred direction of the PA, and, hence the lack of a clear systematic distribution of redshifts.

\begin{figure}
    \centering
  \includegraphics[width=\linewidth, keepaspectratio]{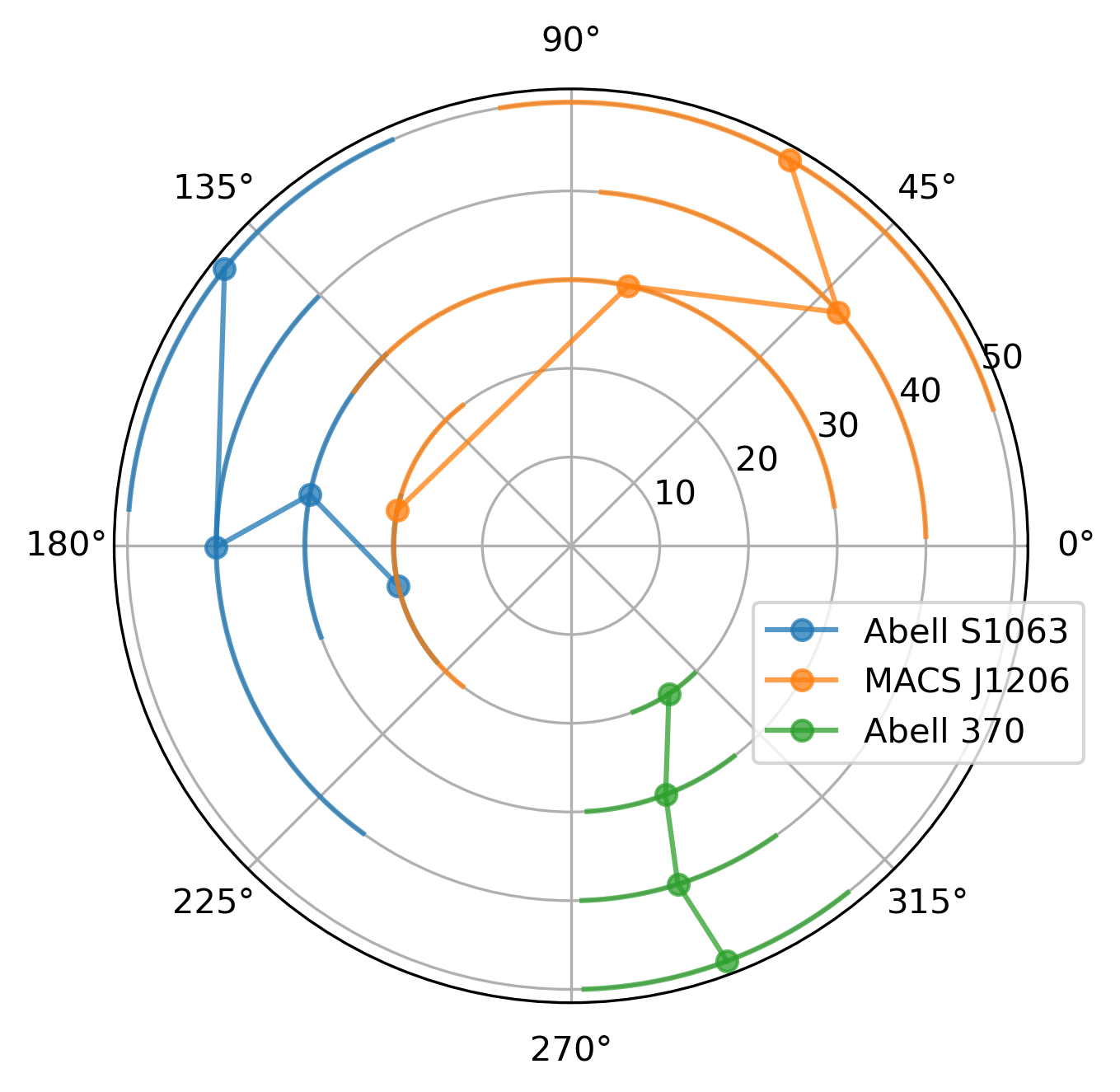}
  \caption{Medians of the fitted PAs for the four smallest values of $N$ (the radial coordinate) for the three clusters in the bias-test sample. The coloured arcs represent the intervals between the 16$^\mathrm{th}$ and 84$^\mathrm{th}$ percentiles of PAs.}\label{fig:rich_clusters_core_PA}
\end{figure}

We can try to use this result to strengthen the evidence for rotation in the core of Abell 370 that we argued for when looking at the deviation from the model of the bias in Fig.\;\ref{fig:Bias_clusters}. Using the same bootstrapping technique we introduced in Sect.\;\ref{sect:Bias testing with radial cuts}, we plot in Fig.\;\ref{fig:rich_clusters_core_PA} the median values of the fitted PAs in the four innermost radii (corresponding to $N=20,30,40,50$, where the deviation in Fig.\;\ref{fig:Bias_clusters} is most noticeable). To assess the stability of these PA measurements, we also show the interval between the 16$^\mathrm{th}$ and 84$^\mathrm{th}$ percentiles for each $N$, providing a measure of their statistical uncertainty. 
Interestingly, A370 PAs vary sensibly less when compared with the other two clusters, with a maximum angular separation between the medians of $16\degr$, while for Abell S1063 such separation is $52\degr$ and for MACS J1206 is $127\degr$. We can further characterise this fact by considering the minimum and maximum $P_{84}-P_{16}$ interval for the three clusters, which are respectively $57\degr$ ($N=20$) and $98\degr$ ($N=40$) for Abell S1063, $81\degr$ ($N=50$) and $136\degr$ ($N=30$) for MACS J1206, and $24\degr$ ($N=20$) and $36\degr$ ($N=50$) for A370.
Its narrow interval of PAs apparently confirms that the innermost region of the galactic component of Abell 370 has indeed a non-negligible amount of systematic rotation.

\section{Conclusions}\label{ConclusionsSection}
The first objective of this study was to investigate the presence of systematic rotation of the galactic component in a sample of 17 rich and nearby ($z<0.1$) Abell clusters. To achieve this, we applied a standard algorithm to extract the rotation parameters ($v_\mathrm{rot}$, PA) by fitting the difference in the average line-of-sight velocity of galaxies, $\Delta v(\theta)$, between the two regions defined by an axis at position angle $\theta$, as a function of $\theta$ itself. After rejecting the clusters that exhibit clear evidence of subclustering in the optical data, we are left with eight clusters that show a sufficiently high rotational support over velocity dispersion ($v_\mathrm{rot}/\sigma\ga0.15$). Three of them, A1650 ($v_\mathrm{rot}\approx176$ km/s, PA $\approx 109\degr$, $v_\mathrm{rot}/\sigma=0.23$), A2029 ($v_\mathrm{rot}\approx 229$ km/s, PA $\approx58\degr$, $v_\mathrm{rot}/\sigma =0.18$) and A2199 ($v_\mathrm{rot}\approx 154$ km/s, PA $\approx78\degr$, $v_\mathrm{rot}/\sigma =0.21$) do not show any evidence of a recent merging event in X-ray data and previous literature considers them to be typical, cool-core, relaxed galaxy clusters.

We also investigated whether the number of kinematic data points has an effect on the determination of the rotation parameters. Naively, one might assume that when using a reduced sample of member galaxies, the fitted rotational amplitude and axis position angle would still provide accurate, though less precise, estimates. Our findings indicate that this is not always the case.
Using the full set of eight tentatively rotating clusters with a sufficiently regular optical morphology, we find a clear tendency to overestimate the amplitude of the rotation signal when the dataset is reduced by multiple random resampling at increasingly smaller values of $N$ (see Fig. \ref{fig:Bias_remaining}).
The effect also depends on the rotational velocity, since the faster the rotation the smaller the number of galaxies required to minimise the bias. To quantify this relation, we introduced a simple analytical model (see Eq. \ref{eq:ratio_as_N} and Appendix \ref{app:bias_derivation}) and used the eight clusters to find the best-fit value for the free correlation parameter $\rho$ associated with the model. The optimal value for this parameter was extracted by least square minimisation and was found to be $\rho_\mathrm{opt}\approx -0.79$.

The final model was used to determine the minimum value of $N$ required to limit the fractional increment $\varepsilon$ of the measured ratio  over its limit value $v_\mathrm{rot}/\sigma$ (see Eq. \ref{eq:increment}). Our results indicate that a reliable ($\varepsilon \la 0.15$) detection of non-negligible rotation ($v_\mathrm{rot}/\sigma\ga 0.15$) should utilise at least 120 {data points with measured kinematics.}.

Finally, we tested whether our model could account for the increase in the $v_\mathrm{rot}/\sigma$ parameter when focussing on the central regions of the clusters. To do this, we used high quality MUSE data from three rich clusters and ran the rotation detection algorithm described above on subsets of galaxies obtained from concentric radial cuts. While for two of them (Abell S1063 and MACS J1206) the data points remain fairly centred around the analytical model curve of the bias, the innermost values of Abell 370 deviate significantly from what the model predicts (Fig.\;\ref{fig:Bias_clusters}). Thus, we argue that this deviation confirms the presence of a significant rotation of the core of Abell 370, supported by the fact that the rotation axis PAs for these regions in this cluster are contained within an interval  much narrower than those for the other two clusters (see Fig.\;\ref{fig:rich_clusters_core_PA}).

\begin{acknowledgements}
The authors thank the anonymous referee for useful comments that
improved the presentation of the work. We acknowledge financial support through grants MIUR2017 WSCC32 and MIUR2020 SKSTHZ. We thank P. Rosati and A. Mercurio for the spectroscopic datasets of Abell S1063 and MACS J1206 and for useful discussions.
\end{acknowledgements}

\bibpunct{(}{)}{;}{a}{}{,} 
\bibliographystyle{aa}
\bibliography{bibliography}

\begin{thebibliography}{57}
\expandafter\ifx\csname natexlab\endcsname\relax\def\natexlab#1{#1}\fi

\bibitem[{{Abell} {et~al.}(1989){Abell}, {Corwin}, \& {Olowin}}]{Abell1989}
{Abell}, G.~O., {Corwin}, Harold~G., J., \& {Olowin}, R.~P. 1989, \apjs, 70, 1

\bibitem[{{Allen} {et~al.}(2011){Allen}, {Evrard}, \& {Mantz}}]{Allen2011}
{Allen}, S.~W., {Evrard}, A.~E., \& {Mantz}, A.~B. 2011, \araa, 49, 409

\bibitem[{{Almeida} {et~al.}(2023){Almeida}, {Anderson}, {Argudo-Fern{\'a}ndez}, {Badenes}, {Barger}, {Barrera-Ballesteros}, {Bender}, {Benitez}, {Besser}, {Bird}, {Bizyaev}, {Blanton}, {Bochanski}, {Bovy}, {Brandt}, {Brownstein}, {Buchner}, {Bulbul}, {Burchett}, {Cano D{\'\i}az}, {Carlberg}, {Casey}, {Chandra}, {Cherinka}, {Chiappini}, {Coker}, {Comparat}, {Conroy}, {Contardo}, {Cortes}, {Covey}, {Crane}, {Cunha}, {Dabbieri}, {Davidson}, {Davis}, {de Andrade Queiroz}, {De Lee}, {M{\'e}ndez Delgado}, {Demasi}, {Di Mille}, {Donor}, {Dow}, {Dwelly}, {Eracleous}, {Eriksen}, {Fan}, {Farr}, {Frederick}, {Fries}, {Frinchaboy}, {G{\"a}nsicke}, {Ge}, {Gonz{\'a}lez {\'A}vila}, {Grabowski}, {Grier}, {Guiglion}, {Gupta}, {Hall}, {Hawkins}, {Hayes}, {Hermes}, {Hern{\'a}ndez-Garc{\'\i}a}, {Hogg}, {Holtzman}, {Ibarra-Medel}, {Ji}, {Jofre}, {Johnson}, {Jones}, {Kinemuchi}, {Kluge}, {Koekemoer}, {Kollmeier}, {Kounkel}, {Krishnarao}, {Krumpe}, {Lacerna}, {Lago}, {Laporte}, {Liu}, {Liu}, {Liu}, {Lopes}, {Macktoobian},
  {Majewski}, {Malanushenko}, {Maoz}, {Masseron}, {Masters}, {Matijevic}, {McBride}, {Medan}, {Merloni}, {Morrison}, {Myers}, {M{\'e}sz{\'a}ros}, {Negrete}, {Nidever}, {Nitschelm}, {Oravetz}, {Oravetz}, {Pan}, {Peng}, {Pinsonneault}, {Pogge}, {Qiu}, {Ramirez}, {Rix}, {Fern{\'a}ndez Rosso}, {Runnoe}, {Salvato}, {Sanchez}, {Santana}, {Saydjari}, {Sayres}, {Schlaufman}, {Schneider}, {Schwope}, {Serna}, {Shen}, {Sobeck}, {Song}, {Souto}, {Spoo}, {Stassun}, {Steinmetz}, {Straumit}, {Stringfellow}, {S{\'a}nchez-Gallego}, {Taghizadeh-Popp}, {Tayar}, {Thakar}, {Tissera}, {Tkachenko}, {Hernandez Toledo}, {Trakhtenbrot}, {Fern{\'a}ndez-Trincado}, {Troup}, {Trump}, {Tuttle}, {Ulloa}, {Vazquez-Mata}, {Vera Alfaro}, {Villanova}, {Wachter}, {Weijmans}, {Wheeler}, {Wilson}, {Wojno}, {Wolf}, {Xue}, {Ybarra}, {Zari}, \& {Zasowski}}]{Almeida2023}
{Almeida}, A., {Anderson}, S.~F., {Argudo-Fern{\'a}ndez}, M., {et~al.} 2023, \apjs, 267, 44

\bibitem[{{Bacon} {et~al.}(2010){Bacon}, {Accardo}, {Adjali}, {Anwand}, {Bauer}, {Biswas}, {Blaizot}, {Boudon}, {Brau-Nogue}, {Brinchmann}, {Caillier}, {Capoani}, {Carollo}, {Contini}, {Couderc}, {Daguis{\'e}}, {Deiries}, {Delabre}, {Dreizler}, {Dubois}, {Dupieux}, {Dupuy}, {Emsellem}, {Fechner}, {Fleischmann}, {Fran{\c{c}}ois}, {Gallou}, {Gharsa}, {Glindemann}, {Gojak}, {Guiderdoni}, {Hansali}, {Hahn}, {Jarno}, {Kelz}, {Koehler}, {Kosmalski}, {Laurent}, {Le Floch}, {Lilly}, {Lizon}, {Loupias}, {Manescau}, {Monstein}, {Nicklas}, {Olaya}, {Pares}, {Pasquini}, {P{\'e}contal-Rousset}, {Pell{\'o}}, {Petit}, {Popow}, {Reiss}, {Remillieux}, {Renault}, {Roth}, {Rupprecht}, {Serre}, {Schaye}, {Soucail}, {Steinmetz}, {Streicher}, {Stuik}, {Valentin}, {Vernet}, {Weilbacher}, {Wisotzki}, \& {Yerle}}]{Bacon2010}
{Bacon}, R., {Accardo}, M., {Adjali}, L., {et~al.} 2010, in Society of Photo-Optical Instrumentation Engineers (SPIE) Conference Series, Vol. 7735, Ground-based and Airborne Instrumentation for Astronomy III, ed. I.~S. {McLean}, S.~K. {Ramsay}, \& H.~{Takami}, 773508

\bibitem[{{Bahcall}(1977)}]{Bahcall1977}
{Bahcall}, N.~A. 1977, \araa, 15, 505

\bibitem[{{Bartalesi} {et~al.}(2024){Bartalesi}, {Ettori}, \& {Nipoti}}]{Bartalesi2024}
{Bartalesi}, T., {Ettori}, S., \& {Nipoti}, C. 2024, \aap, 682, A31

\bibitem[{{Bartalesi} {et~al.}(2025){Bartalesi}, {Ettori}, \& {Nipoti}}]{Bartalesi2025}
{Bartalesi}, T., {Ettori}, S., \& {Nipoti}, C. 2025, \aap, 697, A17

\bibitem[{{Bianchini} {et~al.}(2013){Bianchini}, {Varri}, {Bertin}, \& {Zocchi}}]{Bianchini2013}
{Bianchini}, P., {Varri}, A.~L., {Bertin}, G., \& {Zocchi}, A. 2013, \apj, 772, 67

\bibitem[{{Bianconi} {et~al.}(2013){Bianconi}, {Ettori}, \& {Nipoti}}]{Bianconi2013}
{Bianconi}, M., {Ettori}, S., \& {Nipoti}, C. 2013, \mnras, 434, 1565

\bibitem[{{Buote} \& {Tsai}(1996)}]{BuoteTsai1996}
{Buote}, D.~A. \& {Tsai}, J.~C. 1996, \apj, 458, 27

\bibitem[{{Caminha} {et~al.}(2016){Caminha}, {Grillo}, {Rosati}, {Balestra}, {Karman}, {Lombardi}, {Mercurio}, {Nonino}, {Tozzi}, {Zitrin}, {Biviano}, {Girardi}, {Koekemoer}, {Melchior}, {Meneghetti}, {Munari}, {Suyu}, {Umetsu}, {Annunziatella}, {Borgani}, {Broadhurst}, {Caputi}, {Coe}, {Delgado-Correal}, {Ettori}, {Fritz}, {Frye}, {Gobat}, {Maier}, {Monna}, {Postman}, {Sartoris}, {Seitz}, {Vanzella}, \& {Ziegler}}]{Caminha2016}
{Caminha}, G.~B., {Grillo}, C., {Rosati}, P., {et~al.} 2016, \aap, 587, A80

\bibitem[{{Caminha} {et~al.}(2017){Caminha}, {Grillo}, {Rosati}, {Meneghetti}, {Mercurio}, {Ettori}, {Balestra}, {Biviano}, {Umetsu}, {Vanzella}, {Annunziatella}, {Bonamigo}, {Delgado-Correal}, {Girardi}, {Lombardi}, {Nonino}, {Sartoris}, {Tozzi}, {Bartelmann}, {Bradley}, {Caputi}, {Coe}, {Ford}, {Fritz}, {Gobat}, {Postman}, {Seitz}, \& {Zitrin}}]{Caminha2017}
{Caminha}, G.~B., {Grillo}, C., {Rosati}, P., {et~al.} 2017, \aap, 607, A93

\bibitem[{{Carlberg} {et~al.}(1997){Carlberg}, {Yee}, {Ellingson}, {Morris}, {Abraham}, {Gravel}, {Pritchet}, {Smecker-Hane}, {Hartwick}, {Hesser}, {Hutchings}, \& {Oke}}]{Carlberg1997}
{Carlberg}, R.~G., {Yee}, H.~K.~C., {Ellingson}, E., {et~al.} 1997, \apjl, 476, L7

\bibitem[{{Cote} {et~al.}(1995){Cote}, {Welch}, {Fischer}, \& {Gebhardt}}]{Cote1995}
{Cote}, P., {Welch}, D.~L., {Fischer}, P., \& {Gebhardt}, K. 1995, \apj, 454, 788

\bibitem[{{D'Eugenio} {et~al.}(2015){D'Eugenio}, {Houghton}, {Davies}, \& {Dalla Bont{\`a}}}]{DEugenio2015}
{D'Eugenio}, F., {Houghton}, R.~C.~W., {Davies}, R.~L., \& {Dalla Bont{\`a}}, E. 2015, \mnras, 451, 827

\bibitem[{{Djorgovski} \& {Meylan}(1994)}]{DjorgovskiMeylan1994}
{Djorgovski}, S. \& {Meylan}, G. 1994, \aj, 108, 1292

\bibitem[{{Dressler}(1981)}]{Dressler1981}
{Dressler}, A. 1981, \apj, 243, 26

\bibitem[{{Ettori} {et~al.}(2013){Ettori}, {Donnarumma}, {Pointecouteau}, {Reiprich}, {Giodini}, {Lovisari}, \& {Schmidt}}]{Ettori2013}
{Ettori}, S., {Donnarumma}, A., {Pointecouteau}, E., {et~al.} 2013, \ssr, 177, 119

\bibitem[{{Ettori} \& {Eckert}(2022)}]{Ettori2022}
{Ettori}, S. \& {Eckert}, D. 2022, \aap, 657, L1

\bibitem[{{Ferrami} {et~al.}(2023){Ferrami}, {Bertin}, {Grillo}, {Mercurio}, \& {Rosati}}]{Ferrami2023}
{Ferrami}, G., {Bertin}, G., {Grillo}, C., {Mercurio}, A., \& {Rosati}, P. 2023, \aap, 676, A66

\bibitem[{{Gianfagna} {et~al.}(2021){Gianfagna}, {De Petris}, {Yepes}, {De Luca}, {Sembolini}, {Cui}, {Biffi}, {K{\'e}ruzor{\'e}}, {Mac{\'\i}as-P{\'e}rez}, {Mayet}, {Perotto}, {Rasia}, \& {Ruppin}}]{Gianfagna2021}
{Gianfagna}, G., {De Petris}, M., {Yepes}, G., {et~al.} 2021, \mnras, 502, 5115

\bibitem[{{Govoni} {et~al.}(2009){Govoni}, {Murgia}, {Markevitch}, {Feretti}, {Giovannini}, {Taylor}, \& {Carretti}}]{Govoni2009}
{Govoni}, F., {Murgia}, M., {Markevitch}, M., {et~al.} 2009, \aap, 499, 371

\bibitem[{{Hwang} \& {Lee}(2007)}]{HwangLee2007}
{Hwang}, H.~S. \& {Lee}, M.~G. 2007, \apj, 662, 236

\bibitem[{{Jeans}(1915)}]{Jeans1915}
{Jeans}, J.~H. 1915, \mnras, 76, 70

\bibitem[{{Kacharov} {et~al.}(2014){Kacharov}, {Bianchini}, {Koch}, {Frank}, {Martin}, {van de Ven}, {Puzia}, {McDonald}, {Johnson}, \& {Zijlstra}}]{Kacharov2014}
{Kacharov}, N., {Bianchini}, P., {Koch}, A., {et~al.} 2014, \aap, 567, A69

\bibitem[{{Kalinkov} {et~al.}(2005){Kalinkov}, {Valchanov}, {Valtchanov}, {Kuneva}, \& {Dissanska}}]{Kalinkov2005}
{Kalinkov}, M., {Valchanov}, T., {Valtchanov}, I., {Kuneva}, I., \& {Dissanska}, M. 2005, \mnras, 359, 1491

\bibitem[{{Kopylov} \& {Kopylova}(2007)}]{KopylovKopylova2007}
{Kopylov}, A.~I. \& {Kopylova}, F.~G. 2007, Astrophysical Bulletin, 62, 311

\bibitem[{{Lagattuta} {et~al.}(2022){Lagattuta}, {Richard}, {Bauer}, {Cerny}, {Claeyssens}, {Guaita}, {Jauzac}, {Jeanneau}, {Koekemoer}, {Mahler}, {Prieto Lyon}, {Acebron}, {Meneghetti}, {Niemiec}, {Zitrin}, {Bianconi}, {Connor}, {Cen}, {Edge}, {Faisst}, {Limousin}, {Massey}, {Sereno}, {Sharon}, \& {Weaver}}]{Lagattuta2022}
{Lagattuta}, D.~J., {Richard}, J., {Bauer}, F.~E., {et~al.} 2022, \mnras, 514, 497

\bibitem[{{Lagattuta} {et~al.}(2019){Lagattuta}, {Richard}, {Bauer}, {Cl{\'e}ment}, {Mahler}, {Soucail}, {Carton}, {Kneib}, {Laporte}, {Martinez}, {Patr{\'\i}cio}, {Payne}, {Pell{\'o}}, {Schmidt}, \& {de la Vieuville}}]{Lagattuta2019}
{Lagattuta}, D.~J., {Richard}, J., {Bauer}, F.~E., {et~al.} 2019, \mnras, 485, 3738

\bibitem[{{Landau} \& {Lifshitz}(1980)}]{LandauLifshitz1980}
{Landau}, L.~D. \& {Lifshitz}, E.~M. 1980, {Statistical physics. Pt.1, Pt.2}

\bibitem[{{Le F{\`e}vre} {et~al.}(2003){Le F{\`e}vre}, {Saisse}, {Mancini}, {Brau-Nogue}, {Caputi}, {Castinel}, {D'Odorico}, {Garilli}, {Kissler-Patig}, {Lucuix}, {Mancini}, {Pauget}, {Sciarretta}, {Scodeggio}, {Tresse}, \& {Vettolani}}]{LeFevre2003}
{Le F{\`e}vre}, O., {Saisse}, M., {Mancini}, D., {et~al.} 2003, in Society of Photo-Optical Instrumentation Engineers (SPIE) Conference Series, Vol. 4841, Instrument Design and Performance for Optical/Infrared Ground-based Telescopes, ed. M.~{Iye} \& A.~F.~M. {Moorwood}, 1670--1681

\bibitem[{{Leanza} {et~al.}(2023){Leanza}, {Pallanca}, {Ferraro}, {Lanzoni}, {Dalessandro}, {Cadelano}, {Vesperini}, {Origlia}, {Mucciarelli}, \& {Valenti}}]{Leanza2023}
{Leanza}, S., {Pallanca}, C., {Ferraro}, F.~R., {et~al.} 2023, \apj, 944, 162

\bibitem[{{Leanza} {et~al.}(2022){Leanza}, {Pallanca}, {Ferraro}, {Lanzoni}, {Dalessandro}, {Origlia}, {Mucciarelli}, {Valenti}, {Tiongco}, {Varri}, \& {Vesperini}}]{Leanza2022}
{Leanza}, S., {Pallanca}, C., {Ferraro}, F.~R., {et~al.} 2022, \apj, 929, 186

\bibitem[{{Liu} \& {Tozzi}(2019)}]{LiuTozzi2019}
{Liu}, A. \& {Tozzi}, P. 2019, \mnras, 485, 3909

\bibitem[{{Lotz} {et~al.}(2017){Lotz}, {Koekemoer}, {Coe}, {Grogin}, {Capak}, {Mack}, {Anderson}, {Avila}, {Barker}, {Borncamp}, {Brammer}, {Durbin}, {Gunning}, {Hilbert}, {Jenkner}, {Khandrika}, {Levay}, {Lucas}, {MacKenty}, {Ogaz}, {Porterfield}, {Reid}, {Robberto}, {Royle}, {Smith}, {Storrie-Lombardi}, {Sunnquist}, {Surace}, {Taylor}, {Williams}, {Bullock}, {Dickinson}, {Finkelstein}, {Natarajan}, {Richard}, {Robertson}, {Tumlinson}, {Zitrin}, {Flanagan}, {Sembach}, {Soifer}, \& {Mountain}}]{Lotz2017}
{Lotz}, J.~M., {Koekemoer}, A., {Coe}, D., {et~al.} 2017, \apj, 837, 97

\bibitem[{{Lynden-Bell}(1967)}]{Lynden-Bell1967}
{Lynden-Bell}, D. 1967, \mnras, 136, 101

\bibitem[{{Markevitch} {et~al.}(1999){Markevitch}, {Vikhlinin}, {Forman}, \& {Sarazin}}]{Markevitch1999}
{Markevitch}, M., {Vikhlinin}, A., {Forman}, W.~R., \& {Sarazin}, C.~L. 1999, \apj, 527, 545

\bibitem[{{Materne} \& {Hopp}(1983)}]{MaterneHopp1983}
{Materne}, J. \& {Hopp}, U. 1983, \aap, 124, L13

\bibitem[{{Metropolis} {et~al.}(1953){Metropolis}, {Rosenbluth}, {Rosenbluth}, {Teller}, \& {Teller}}]{Metropolis1953}
{Metropolis}, N., {Rosenbluth}, A.~W., {Rosenbluth}, M.~N., {Teller}, A.~H., \& {Teller}, E. 1953, \jcp, 21, 1087

\bibitem[{{Monteiro-Oliveira} {et~al.}(2022){Monteiro-Oliveira}, {Morell}, {Sampaio}, {Ribeiro}, \& {de Carvalho}}]{Monteiro-Oliveira2022}
{Monteiro-Oliveira}, R., {Morell}, D.~F., {Sampaio}, V.~M., {Ribeiro}, A.~L.~B., \& {de Carvalho}, R.~R. 2022, \mnras, 509, 3470

\bibitem[{{Navarro} {et~al.}(1997){Navarro}, {Frenk}, \& {White}}]{Navarro1997}
{Navarro}, J.~F., {Frenk}, C.~S., \& {White}, S. D.~M. 1997, \apj, 490, 493

\bibitem[{{Oegerle} \& {Hill}(1992)}]{OegerleHill1992}
{Oegerle}, W.~R. \& {Hill}, J.~M. 1992, \aj, 104, 2078

\bibitem[{{Peebles}(1969)}]{Peebles1969}
{Peebles}, P.~J.~E. 1969, \apj, 155, 393

\bibitem[{{Petralia} {et~al.}(2024){Petralia}, {Minniti}, {Fern{\'a}ndez-Trincado}, {Lane}, \& {Schiavon}}]{Petralia2024}
{Petralia}, I., {Minniti}, D., {Fern{\'a}ndez-Trincado}, J.~G., {Lane}, R.~R., \& {Schiavon}, R.~P. 2024, \aap, 688, A92

\bibitem[{{Planck Collaboration} {et~al.}(2020){Planck Collaboration}, {Aghanim}, {Akrami}, {Ashdown}, {Aumont}, {Baccigalupi}, {Ballardini}, {Banday}, {Barreiro}, {Bartolo}, {Basak}, {Battye}, {Benabed}, {Bernard}, {Bersanelli}, {Bielewicz}, {Bock}, {Bond}, {Borrill}, {Bouchet}, {Boulanger}, {Bucher}, {Burigana}, {Butler}, {Calabrese}, {Cardoso}, {Carron}, {Challinor}, {Chiang}, {Chluba}, {Colombo}, {Combet}, {Contreras}, {Crill}, {Cuttaia}, {de Bernardis}, {de Zotti}, {Delabrouille}, {Delouis}, {Di Valentino}, {Diego}, {Dor{\'e}}, {Douspis}, {Ducout}, {Dupac}, {Dusini}, {Efstathiou}, {Elsner}, {En{\ss}lin}, {Eriksen}, {Fantaye}, {Farhang}, {Fergusson}, {Fernandez-Cobos}, {Finelli}, {Forastieri}, {Frailis}, {Fraisse}, {Franceschi}, {Frolov}, {Galeotta}, {Galli}, {Ganga}, {G{\'e}nova-Santos}, {Gerbino}, {Ghosh}, {Gonz{\'a}lez-Nuevo}, {G{\'o}rski}, {Gratton}, {Gruppuso}, {Gudmundsson}, {Hamann}, {Handley}, {Hansen}, {Herranz}, {Hildebrandt}, {Hivon}, {Huang}, {Jaffe}, {Jones}, {Karakci}, {Keih{\"a}nen},
  {Keskitalo}, {Kiiveri}, {Kim}, {Kisner}, {Knox}, {Krachmalnicoff}, {Kunz}, {Kurki-Suonio}, {Lagache}, {Lamarre}, {Lasenby}, {Lattanzi}, {Lawrence}, {Le Jeune}, {Lemos}, {Lesgourgues}, {Levrier}, {Lewis}, {Liguori}, {Lilje}, {Lilley}, {Lindholm}, {L{\'o}pez-Caniego}, {Lubin}, {Ma}, {Mac{\'\i}as-P{\'e}rez}, {Maggio}, {Maino}, {Mandolesi}, {Mangilli}, {Marcos-Caballero}, {Maris}, {Martin}, {Martinelli}, {Mart{\'\i}nez-Gonz{\'a}lez}, {Matarrese}, {Mauri}, {McEwen}, {Meinhold}, {Melchiorri}, {Mennella}, {Migliaccio}, {Millea}, {Mitra}, {Miville-Desch{\^e}nes}, {Molinari}, {Montier}, {Morgante}, {Moss}, {Natoli}, {N{\o}rgaard-Nielsen}, {Pagano}, {Paoletti}, {Partridge}, {Patanchon}, {Peiris}, {Perrotta}, {Pettorino}, {Piacentini}, {Polastri}, {Polenta}, {Puget}, {Rachen}, {Reinecke}, {Remazeilles}, {Renzi}, {Rocha}, {Rosset}, {Roudier}, {Rubi{\~n}o-Mart{\'\i}n}, {Ruiz-Granados}, {Salvati}, {Sandri}, {Savelainen}, {Scott}, {Shellard}, {Sirignano}, {Sirri}, {Spencer}, {Sunyaev}, {Suur-Uski}, {Tauber}, {Tavagnacco},
  {Tenti}, {Toffolatti}, {Tomasi}, {Trombetti}, {Valenziano}, {Valiviita}, {Van Tent}, {Vibert}, {Vielva}, {Villa}, {Vittorio}, {Wandelt}, {Wehus}, {White}, {White}, {Zacchei}, \& {Zonca}}]{PlanckCollabo2020}
{Planck Collaboration}, {Aghanim}, N., {Akrami}, Y., {et~al.} 2020, \aap, 641, A6

\bibitem[{{Postman} {et~al.}(2012){Postman}, {Coe}, {Ben{\'\i}tez}, {Bradley}, {Broadhurst}, {Donahue}, {Ford}, {Graur}, {Graves}, {Jouvel}, {Koekemoer}, {Lemze}, {Medezinski}, {Molino}, {Moustakas}, {Ogaz}, {Riess}, {Rodney}, {Rosati}, {Umetsu}, {Zheng}, {Zitrin}, {Bartelmann}, {Bouwens}, {Czakon}, {Golwala}, {Host}, {Infante}, {Jha}, {Jimenez-Teja}, {Kelson}, {Lahav}, {Lazkoz}, {Maoz}, {McCully}, {Melchior}, {Meneghetti}, {Merten}, {Moustakas}, {Nonino}, {Patel}, {Reg{\"o}s}, {Sayers}, {Seitz}, \& {Van der Wel}}]{Postman2012}
{Postman}, M., {Coe}, D., {Ben{\'\i}tez}, N., {et~al.} 2012, \apjs, 199, 25

\bibitem[{{Reiprich} \& {B{\"o}hringer}(2002)}]{Reiprich2002}
{Reiprich}, T.~H. \& {B{\"o}hringer}, H. 2002, \apj, 567, 716

\bibitem[{{Ricker}(1998)}]{Ricker1998}
{Ricker}, P.~M. 1998, \apj, 496, 670

\bibitem[{{Roettiger} \& {Flores}(2000)}]{RoettigerFlores2000}
{Roettiger}, K. \& {Flores}, R. 2000, \apj, 538, 92

\bibitem[{{Rood} {et~al.}(1972){Rood}, {Page}, {Kintner}, \& {King}}]{Rood1972}
{Rood}, H.~J., {Page}, T.~L., {Kintner}, E.~C., \& {King}, I.~R. 1972, \apj, 175, 627

\bibitem[{{Rosati} {et~al.}(2014){Rosati}, {Balestra}, {Grillo}, {Mercurio}, {Nonino}, {Biviano}, {Girardi}, {Vanzella}, \& {Clash-VLT Team}}]{Rosati2014}
{Rosati}, P., {Balestra}, I., {Grillo}, C., {et~al.} 2014, The Messenger, 158, 48

\bibitem[{{Sarazin}(1988)}]{Sarazin1988}
{Sarazin}, C.~L. 1988, {X-ray emission from clusters of galaxies}

\bibitem[{{Smith}(1936)}]{Smith1936}
{Smith}, S. 1936, \apj, 83, 23

\bibitem[{{Song} {et~al.}(2018){Song}, {Hwang}, {Park}, {Smith}, \& {Einasto}}]{Song2018}
{Song}, H., {Hwang}, H.~S., {Park}, C., {Smith}, R., \& {Einasto}, M. 2018, \apj, 869, 124

\bibitem[{{Steinhardt} {et~al.}(2020){Steinhardt}, {Jauzac}, {Acebron}, {Atek}, {Capak}, {Davidzon}, {Eckert}, {Harvey}, {Koekemoer}, {Lagos}, {Mahler}, {Montes}, {Niemiec}, {Nonino}, {Oesch}, {Richard}, {Rodney}, {Schaller}, {Sharon}, {Strolger}, {Allingham}, {Amara}, {Bah{\'e}}, {B{\oe}hm}, {Bose}, {Bouwens}, {Bradley}, {Brammer}, {Broadhurst}, {Ca{\~n}as}, {Cen}, {Cl{\'e}ment}, {Clowe}, {Coe}, {Connor}, {Darvish}, {Diego}, {Ebeling}, {Edge}, {Egami}, {Ettori}, {Faisst}, {Frye}, {Furtak}, {G{\'o}mez-Guijarro}, {Remolina Gonz{\'a}lez}, {Gonzalez}, {Graur}, {Gruen}, {Harvey}, {Hensley}, {Hovis-Afflerbach}, {Jablonka}, {Jha}, {Jullo}, {Kneib}, {Kokorev}, {Lagattuta}, {Limousin}, {von der Linden}, {Linzer}, {Lopez}, {Magdis}, {Massey}, {Masters}, {Maturi}, {McCully}, {McGee}, {Meneghetti}, {Mobasher}, {Moustakas}, {Murphy}, {Natarajan}, {Neyrinck}, {O'Connor}, {Oguri}, {Pagul}, {Rhodes}, {Rich}, {Robertson}, {Sereno}, {Shan}, {Smith}, {Sneppen}, {Squires}, {Tam}, {Tchernin}, {Toft}, {Umetsu}, {Weaver}, {van
  Weeren}, {Williams}, {Wilson}, {Yan}, \& {Zitrin}}]{Steinhardt2020}
{Steinhardt}, C.~L., {Jauzac}, M., {Acebron}, A., {et~al.} 2020, \apjs, 247, 64

\bibitem[{{Tovmassian}(2015)}]{Tovmassian2015}
{Tovmassian}, H.~M. 2015, Astrophysics, 58, 328

\bibitem[{{Zocchi} {et~al.}(2012){Zocchi}, {Bertin}, \& {Varri}}]{Zocchi2012}
{Zocchi}, A., {Bertin}, G., \& {Varri}, A.~L. 2012, \aap, 539, A65

\end{thebibliography}

\begin{appendix}
\section{Performance of the two rotation detecting algorithms}\label{app:methods_testing}
In order to evaluate the efficiency of the rotation-detecting algorithms, it would be useful to have a number of rotating cluster-projected images of which the kinematics is known. Since this is not available for real clusters, we must refer to simulated images of clusters of galaxies. We adopt a statistical approach to describe the galactic component by means of a continuous, non-negative, phase-space distribution function $f(\Vec{x},\Vec{v})$. Here, we do not aim at constructing a physically justified model for the galactic component, but only at providing a reasonable representation of the distribution of galaxies in phase space when some systematic rotation is present.

Clusters are bound and quasi-stationary structures and, despite its low collisionality \citep[with respect to galaxy-galaxy scattering; see, e.g.,][]{Bahcall1977}, the galactic component may be considered to have reached an equilibrium state, at least in the central regions. This picture provides arguments in favour of using a distribution function that closely resembles the Maxwell-Boltzmann $f(E)=A\exp\left[-aE\right]$, where $E=\frac{v^2}{2}+\Phi$ is the single-galaxy energy per unit mass and $a=\frac{1}{\sigma^2}$, with $\sigma^2$ the velocity dispersion, assumed here to be constant. The simplest way to introduce rotation is to replace the energy per unit mass with the Jacobi integral 
\begin{equation}
    H = E-\Vec{J}\cdot\Vec{\omega}\:,
\end{equation}
where $\Vec{\omega}$ and $\Vec{J}$ have, respectively, the dimensions of an angular velocity and an angular momentum per unit mass. This quantity represents the single-galaxy energy per unit mass in the frame of reference rotating with a constant angular velocity vector $\Vec{\omega}$ \citep[see][]{LandauLifshitz1980}, and, following the Jeans theorem \citep{Jeans1915}, it is a suitable integral of motion in the case of axisymmetric potentials. This leaves us with the following distribution function:
\begin{equation}\label{eq:Jacobi_DF}
    f=A\exp\left[-\frac{1}{\sigma^2}\left(\frac{v^2}{2}-\Vec{J}\cdot\Vec{\omega}+\Phi\right)\right]\;.
\end{equation}
Although rigid rotation is better motivated (at least for the core region of the cluster), to test the accuracy of the rotation-detecting algorithms we decided to use a rotation law of the form $\omega(R)={v_\mathrm{rot}}/{R}$, where $R$ is the distance from the rotation axis, so that the rotational velocity would remain constant. The apparent violation of the Jeans theorem is not an issue in this context, because we are not interested in the evolution of the galactic component and we only wish to check the quality of the algorithms used, in the presence of systematic motions in the galactic component. Finally, to complete the description of the spatial distribution of galaxies, we model the total gravitational potential with a NFW profile \citep{Navarro1997}:
\begin{equation}
    \Phi=-\Phi_0\frac{r_s}{r}\log\left(1+\frac{r}{r_s}\right)\:.
\end{equation}

To produce simulated images of a rotating cluster, we have developed a Monte Carlo simulation that utilises the Metropolis algorithm \citep{Metropolis1953} to sample 250 phase-space points (the three dimensional position and velocity vectors of the galaxies) of 500 clusters, from the simple distribution function in Eq.\;(\ref{eq:Jacobi_DF}). In this way, even when we narrow our field of view to match the scale radius $R$, the number of data points drops to a typical value of the order of 120, which is still large enough to be considered unbiased, regardless of the value of $v_\mathrm{rot}/\sigma$ (see Fig.\;\ref{fig:minN}).

We tested the two most used rotation-detecting methods. The first one is that described in Sect.\;\ref{MethodsSection}: here, for simplicity, we will refer to it as AVD (Average Velocity Difference). The second one was introduced by \citet{HwangLee2007}, here referred to as HL: it employs a fitting procedure for the $i$-th observed galaxy with radial velocity $v_i$ and projected angular position $\theta_i$, using the function
\begin{equation}
    v_{p,i}(v_{\mathrm{rot}}, \theta_{\mathrm{ax}})=v_{\mathrm{rot}}\sin(\theta_i-\theta_{\mathrm{ax}})\:.
\end{equation}
The optimal fitting parameters $v_\mathrm{rot}$ and $\theta_\mathrm{ax}$ are extracted through the least-square minimisation of
\begin{equation}
    \chi^2=\sum_i(v_{p,i}-v_i)^2\:.
\end{equation}

\begin{figure*}
    \centering
  \includegraphics[width=0.75\linewidth, keepaspectratio]{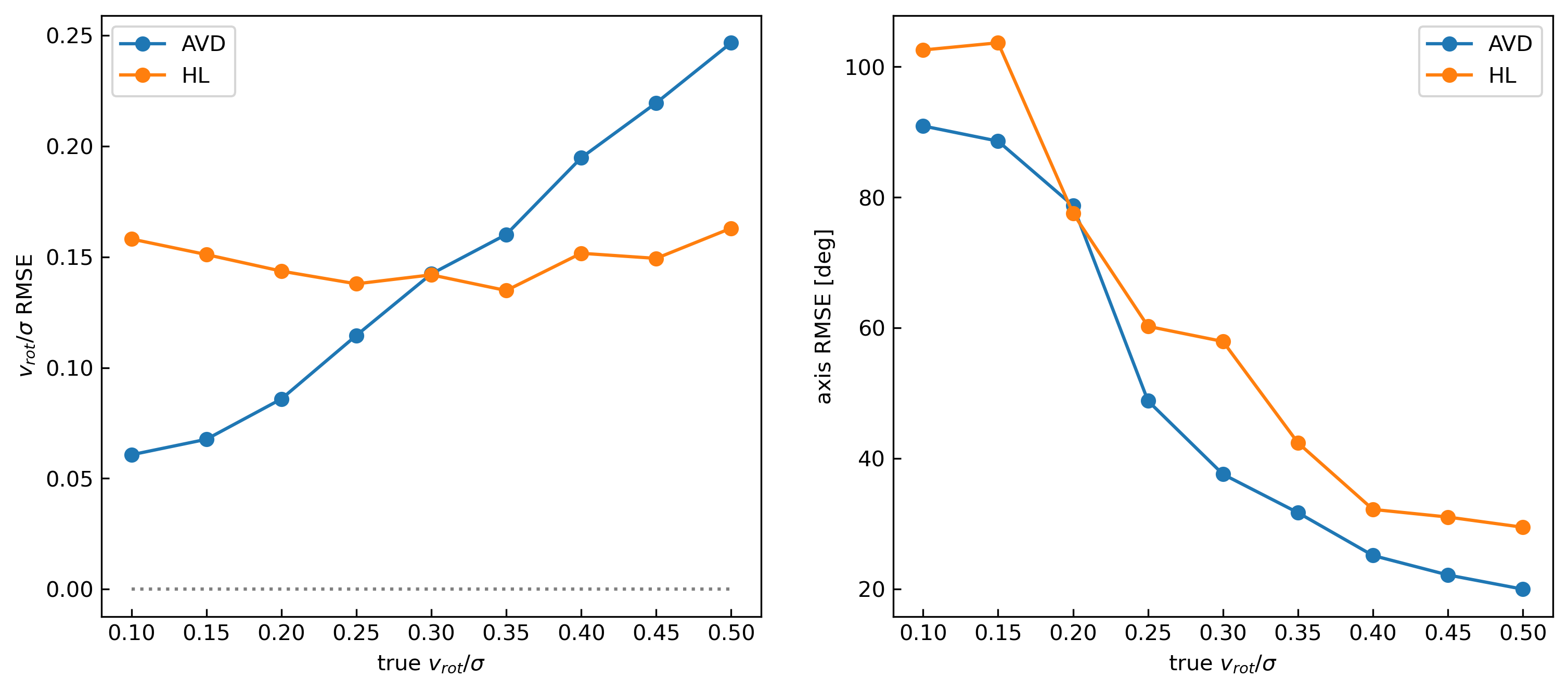}
  \caption{Root mean square errors of $v_\mathrm{rot}/\sigma$ and projected axis position angle extracted with the two methods described in the main text.}\label{fig:Methods_comparison}
\end{figure*}

The results are shown in Fig. \ref{fig:Methods_comparison}.
It is important to note that the performance of the two methods depends on the adopted distribution function: in particular, the results may vary for a different angular velocity law $\omega(R)$, a different gravitational potential and even for a different number of galaxies. For this reason, we argue that the main conclusions of this test are that the AVD method is likely to be more appropriate for slow rotating clusters; for higher rotational velocities, the HL method is better. In addition, the AVD method appears to be more accurate in recovering the correct axis orientation.

\section{Derivation of Equation (\ref{eq:ratio_as_N})}\label{app:bias_derivation}
Let us call the sample of galaxies on one side of a projected axis passing through the cluster centre $A$ and the sample on the other side $B$.
The starting assumptions are that the distributions of velocities of $A$ and $B$ are Gaussian with different mean but with the same variance, which is the same as the variance of the total sample.
This can be written as 
\begin{equation}
v_{A/B} \sim \mathcal{N}\left(\mu_{A/B}, \sigma_{A/B}^2\right) \:.
\end{equation}\label{eq:gaussian_vel}

If the axis that divides the two samples is the rotation axis, then we expect $\mu_{A} = - \mu_{B}$ and $\sigma_{A} = \sigma_{B}$.
Furthermore we know that since we have already subtracted the cluster member velocities from the Hubble flow, the overall distribution of velocities is 
\begin{equation}
v \sim  \mathcal{N}\left(\mu_{A}, \sigma_{A}^2\right) + \mathcal{N}\left(\mu_{B}, \sigma_{B}^2\right) =
        \mathcal{N}\left(0, 2\sigma_{A}^2 (1+\rho)\right) = \mathcal{N}\left(0, \sigma^2 \right)\:,
\end{equation}\label{eq:gaussian_vel_overall}
where $\rho$ is the Pearson correlation coefficient. 
The Pearson correlation coefficient is defined as
\begin{equation}
\begin{split}
\rho & = \frac{\mathrm{Cov}(v_A, v_B)}{\sigma_A \sigma_B} =
\frac{\mathbb{E}\left[(v_A-\mu_A)(v_B-\mu_B)\right]}{\sigma_A \sigma_B}\\
& =\frac{\int_0^{R_{\rm{max}}}\int_0^{R_{\rm{max}}}(v_A(R_A)-\mu)(v_B(R_B)-\mu)\bar\Sigma(R_A, R_B) \dd R_A \dd R_B}{\sigma_A \sigma_B} \:,
\end{split}
\end{equation}\label{eq:Pearson_coeff}
where $\bar\Sigma(R_A, R_B)$ is the joint probability density of drawing two galaxies at $R_A$ and $R_B$ within a maximum radius $R_{\rm{max}}$.
The correlation coefficient is a dimensionless and bounded ($-1 \leq \rho \leq 1$) measure of the covariance between two samples of galaxy line-of-sight velocities drawn from opposite sides of the rotation axis.
The value of $\rho$ incorporates the effect of slight asymmetries in the surface number density and the effect of the random motion component of the member galaxy velocity distribution via the structure of the joint probability density.
We expect the velocity measurements at opposite sides of the rotation axis to be anticorrelated, therefore $-1<\rho<0$ and we can then define $\hat\rho = 1/(1+\rho)$.
We consider $\rho$ to be a summary statistics of the structure and dynamics of the cluster members within a maximum projected radius, and we estimate its value from a fit of a set of clusters in Sect.\;\ref{Rotation detection and model calibration Subsection}. 

Considering now a finite sample of cluster members, the Central Limit Theorem (CLT) leads to
\begin{equation}
\langle v_{A/B} \rangle 
\sim \mathcal{N}\left(\mu_{A/B}, \: \sigma_{A}^2\frac{2}{N}\right)
= \mathcal{N}\left(\mu_{A/B}, \: \sigma^2\frac{\hat\rho}{N}\right) \:,
\end{equation}\label{eq:CLT}
assuming that the cluster contains $N$ galaxies, half in $A$ and half in $B$.

The halved difference between the mean velocities in $A$ and $B$ is distributed as
\begin{equation}\label{eq:diff_gauss}
\frac{\langle v_{A} \rangle - \langle v_{B} \rangle}{2} = \Delta v \sim 
\mathcal{N}\left(v_\mathrm{rot}, \sigma^2\frac{\hat\rho}{2N}\right) \:,
\end{equation}
where $v_\mathrm{rot} = (\mu_{A} - \mu_{B})/2$.

Using the properties of the folded normal distribution, the mean of the absolute value of Eq.\;(\ref{eq:diff_gauss}) is
\begin{equation}
\begin{split}
    \langle | \Delta v |  \rangle &=
    \left\langle \left| 
    \mathcal{N}\left(v_\mathrm{rot}, \sigma^2\frac{\hat\rho}{2N}\right) 
    \right| \right\rangle \\
    &= \sigma \sqrt{\frac{\hat\rho}{\pi N}} 
    \exp\left(-\frac{v_\mathrm{rot}^2}{\sigma^2}\frac{N}{\hat\rho}\right)+
    v_\mathrm{rot}\erf\left(\frac{v_\mathrm{rot}}{\sigma}\sqrt{\frac{N}{\hat\rho}}\right) \:.
\end{split}
\end{equation}

Noting that the standard deviation $\tilde\sigma$ of a finite sample extracted from a Gaussian distribution with variance $\sigma^2$ is
\begin{equation}
\tilde\sigma = \sigma \sqrt{\frac{2}{N-1}}\frac{\Gamma(N/2)}{\Gamma((N-1)/2)} \:,
\end{equation}\label{eq:std_dev_N}
we obtain Eq.\;(\ref{eq:ratio_as_N}).

\section{{Simultaneous fit of $v_\mathrm{rot}/\sigma$ and $\rho$}}\label{app:simult_fit}

\begin{figure}
    \centering
  \includegraphics[width=\linewidth, keepaspectratio]{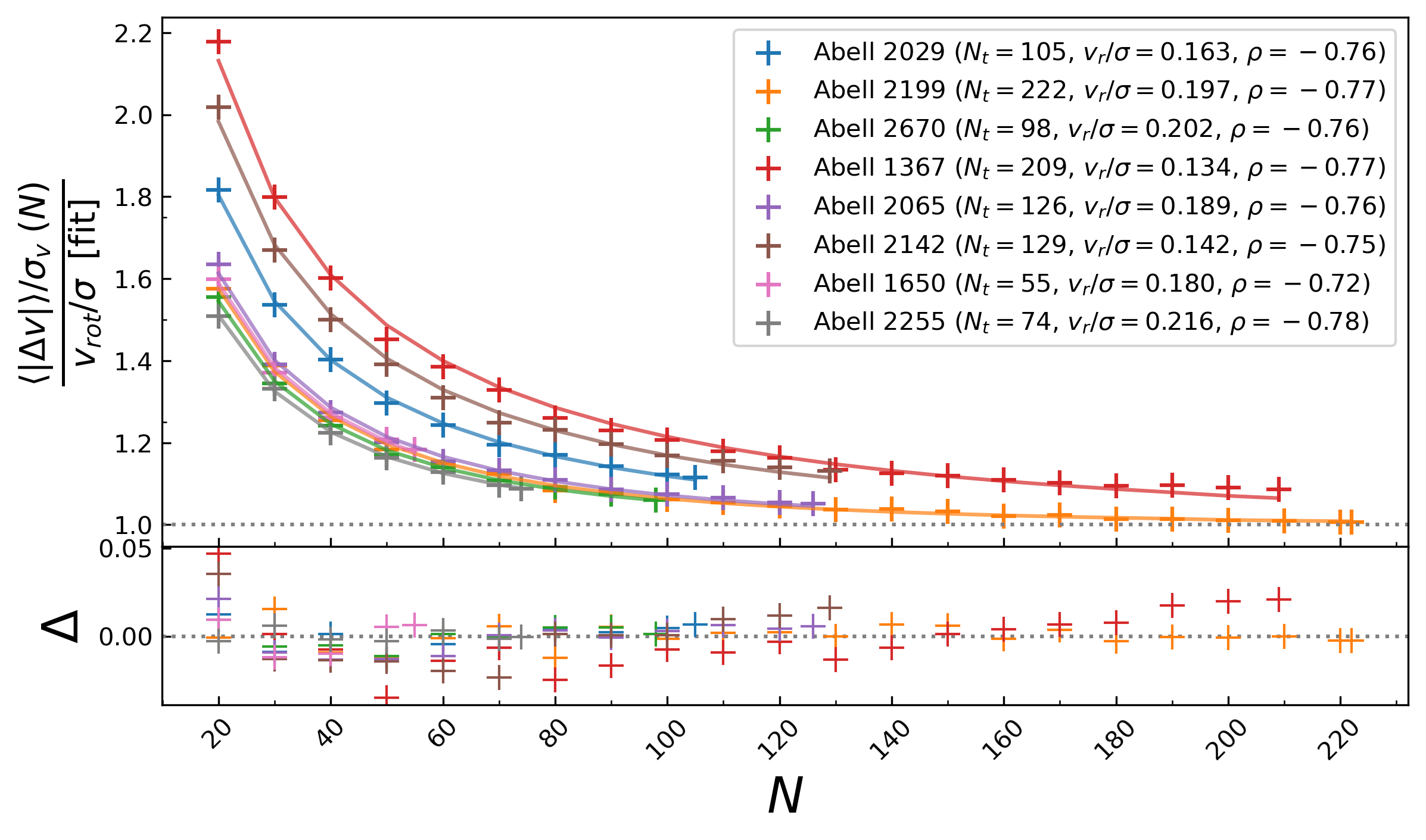}
  \caption{{Low-number bias trends of the eight selected clusters in the paper as a function of the number $N$ of galaxies included in the analysis. The continuous lines are our model best fit curves to the data, where both $v_\mathrm{rot}/\sigma$ and $\rho$ are free parameters. Their fitted values are reported in the legend, alongside the total number of galaxies $N_t=N_\mathrm{tot}$.}}\label{fig:separate_fit}
\end{figure}

A simultaneous fit of both $v_\mathrm{rot}/\sigma$ and $\rho$ allows us to retrieve a new estimate of the limit rotation velocity over velocity dispersion ratio and a distinct correlation coefficient for each cluster, whose value we previously imposed to be a constant. The resulting fits are shown in Fig. \ref{fig:separate_fit}.
Note that we previously constrained the curves to 1 for $N = N_\mathrm{tot}$ because we assumed this was the 'true' value of $v_\mathrm{rot}/\sigma$, obtained using the maximum available statistics. In practice, removing this constraint is equivalent to treating the $N = N_\mathrm{tot}$ case as an approximation of the ideal limit with infinite sample size. Consequently, the values of $|\Delta v| / \sigma_v$ at $N = N_\mathrm{tot}$ (referred to as $v_\mathrm{rot}/\sigma$ in Fig. \ref{fig:Bias_remaining}) are in this case also affected by the low-number bias, and thus we expect them to be slightly larger than the fitted values of $v_\mathrm{rot}/\sigma$ in Fig. \ref{fig:separate_fit}. However, as with any other $N$, the size of the bias in the measured value of $|\Delta v| / \sigma_v$  at $N=N_\mathrm{tot}$ depends both on the fitted $v_\mathrm{rot}/\sigma$ ratio and $N_\mathrm{tot}$. This explains why the limit ratio of a relatively fast-rotating, rich cluster, such as A2199, has barely changed, whereas that of the slower A1367 has decreased more, despite its $N_\mathrm{tot}$ being almost as large as that of A2199.

Although the optimal values of $\rho$ are slightly larger than those obtained using the first method, they are remarkably consistent with each other, with an average value of $-0.759 \pm 0.015$. This confirms that this coefficient remains approximately the same for every cluster. As already mentioned in Appendix \ref{app:bias_derivation}, its numerical value is directly linked to the structural and kinematical properties of the clusters through the joint probability density function $\bar\Sigma(R_A, R_B)$. However, it is not easy to justify and construct a distribution function that can reproduce our estimate of $\rho$, so, for the moment, the the origin of its numerical value remains unclear.

\section{Low-number bias on a sample of globular clusters}\label{app:bias_globular_cluster}
As noted above, the low-number bias is not exclusive to galaxy clusters, but it is of interest in all cases where the search for rotation is carried out using line-of-sight velocities alone.

For instance, this is the case when studying the kinematics of globular clusters for which star proper motions are not known or not included in the analysis. Here, we consider a recent paper by \citet{Petralia2024}, in which they analyse the kinematics of a sample of 21 globular clusters from the APOGEE-2 survey and select the clusters exhibiting signatures of systematic rotation. The principal method they use is essentially identical to the one used in this paper, with a slight change in the notation: $N_*=N_\mathrm{tot}$ and $A_\mathrm{fit}=2|\Delta v|$. Their findings reveal that 21 of the 23 globular clusters included in their sample show a signature of systematic rotation (we should note, however, that they adopt a looser definition of rotating cluster).

The results on the bias discussed in this paper suggest that a reassessment of the fitted rotational velocity of some of the globular clusters in \citet{Petralia2024} might be appropriate. In particular, Table \ref{tab:globular_clusters} highlights the globular clusters that appear most likely to be affected by the presence of the bias. Specifically, our model suggests that the signal observed in the three globular clusters with $|\Delta v|/\sigma_v\leq 0.10$ (NGC 6171, NGC 6273, and NGC 6752) could be largely attributed to the low-number bias.

\begin{table}[h!]
    \centering
    \caption{Rotation parameters for some of the globular clusters in \citet{Petralia2024}.}
    \label{tab:globular_clusters}
    \begin{tabular}{lcccc}
    \hline\hline
    ID    & $N_\mathrm{tot}$    & $\sigma_v$     & $|\Delta v|$      &   $|\Delta v|/\sigma_v$   \\
    & & (km/s) & (km/s) & \\
    \hline
    NGC 0362          &     70    &    8.6     &    $1.79$    &   0.21   \\
    NGC 6171        &    65      &    4.1      &   $0.39$     &   0.09    \\
    NGC 6254        &     87     &    6.3     &    $1.14$     &   0.18   \\
    NGC 6273        &     81     &     12.1     &   $1.19$      &    0.10   \\
    NGC 6388        &    75      &    17.4      &    $3.32$     &    0.19    \\
    NGC 6397        &     187     &    5.5      &    $0.70$     &   0.13    \\
    NGC 6752        &   152       &     7.7     &   $0.52$      &    0.07   \\
    \hline
    \end{tabular}
    \tablefoot{$N_\mathrm{tot}=N_*$ refers to the number of the selected member stars. The third and fourth column display, respectively, the velocity dispersion and the rotation amplitude calculated as $|\Delta v|=A_\mathrm{fit}/2$.}
\end{table}

\end{appendix}
\end{document}